\documentclass[,twocolumn]{aastex631}

\newcommand{\Lagr}{\mathcal{L}}
\usepackage{natbib}
\usepackage{newtxtext,newtxmath}
\usepackage{amsmath}
\usepackage{graphics,graphicx}
\newcommand{\angstrom}{\mbox{\normalfont\AA}}

\usepackage{subcaption}
\usepackage{mathtools}
\usepackage{enumitem}
\usepackage{lipsum}
\setlist[itemize]{leftmargin=*}

\usepackage{floatrow}
\usepackage[outercaption]{sidecap}    
\sidecaptionvpos{figure}{t}

\usepackage[T1]{fontenc}
\usepackage{ae,aecompl}
\pdfoutput=1

\shorttitle{High-z stellar masses can be recovered robustly with JWST photometry}
\shortauthors{R. K. Cochrane et al.}

\begin{document}
\title{High-z stellar masses can be recovered robustly with JWST photometry}
\author[0000-0001-8855-6107]{R. K. Cochrane}
\affiliation{Institute for Astronomy, University of Edinburgh, Royal Observatory, Blackford Hill, Edinburgh, EH9 3HJ, UK}
\affiliation{Department of Astronomy, Columbia University, New York, NY 10027, USA}
\footnote{email: rcochra3@ed.ac.uk}

\author{H. Katz}
\affiliation{Department of Astronomy and Astrophysics, University of Chicago, 5640 S Ellis Avenue, Chicago, IL 60637, USA}

\author[0000-0003-0629-8074]{R. Begley}
\affiliation{Institute for Astronomy, University of Edinburgh, Royal Observatory, Blackford Hill, Edinburgh, EH9 3HJ, UK}

\author[0000-0003-4073-3236]{C. C. Hayward}
\affiliation{Eureka Scientific, Inc., 2452 Delmer Street, Suite 100, Oakland, CA 94602, USA}
\affiliation{Kavli Institute for the Physics and Mathematics of the Universe (WPI), The University of Tokyo Institutes for Advanced Study, The University of Tokyo, Kashiwa, Chiba 277-8583, Japan}
\affiliation{Center for Computational Astrophysics, Flatiron Institute, 162 Fifth Avenue, New York, NY 10010, USA}

\author{P. N. Best}
\affiliation{Institute for Astronomy, University of Edinburgh, Royal Observatory, Blackford Hill, Edinburgh, EH9 3HJ, UK}

\begin{abstract}
\noindent Robust inference of galaxy stellar masses from photometry is crucial for constraints on galaxy assembly across cosmic time. Here, we test a commonly-used Spectral Energy Distribution (SED) fitting code, using simulated galaxies from the S{\sc phinx}$^{20}$ cosmological radiation hydrodynamics simulation, with JWST NIRCam photometry forward-modelled with radiative transfer. Fitting the synthetic photometry with various star formation history models, we show that recovered stellar masses are, encouragingly, generally robust to within a factor of $\sim3$ for galaxies in the range $M_{\star}\sim10^{7}-10^{9}\,\rm{M_{\odot}}$ at $z=5-10$. These results are in stark contrast to recent work claiming that stellar masses can be underestimated by as much as an order of magnitude in these mass and redshift ranges. However, while $>90\%$ of masses are recovered to within $0.5\,\rm{dex}$, there are notable systematic trends, with stellar masses typically overestimated for low-mass galaxies ($M_{\star}\lesssim10^{8}\,\rm{M_{\odot}}$) and slightly underestimated for high-mass galaxies ($M_{\star}\gtrsim10^{9}\,\rm{M_{\odot}}$). We demonstrate that these trends arise due to the SED fitting code poorly modelling the impact of strong emission lines on broadband photometry. These systematic trends, which exist for all star formation history parametrisations tested, have a tilting effect on the inferred stellar mass function, with number densities of massive galaxies underestimated (particularly at the lowest redshifts studied) and number densities of lower-mass galaxies typically overestimated. Overall, this work suggests that we should be optimistic about our ability to infer the masses of high-$z$ galaxies observed with JWST (notwithstanding contamination from AGN) but careful when modelling the impact of strong emission lines on broadband photometry. 
\end{abstract}
\section{Introduction}
Observational constraints on the assembly of stellar mass are foundational for our understanding of galaxy formation and evolution. Large spectroscopic and photometric surveys have enabled constraints on the evolving stellar mass function (SMF) from $z\sim0$ out to $z\sim8-9$ \citep[][among many others]{Cole2001,Ilbert2009,Ilbert2013,Marchesini2009,Baldry2012,Moutard2016,Davidzon2017,Leja2020,Adams2021,McLeod2021,Weaver2023}, with the most distant measurements made using deep Hubble Space Telescope ({\it{HST}}) imaging, which provided rest-frame ultraviolet (UV)-selected samples \citep[e.g.][]{Duncan2014,Grazian2015,Song2016,Bhatawdekar2019,Kikuchihara2020}. JWST is providing a more complete census of stellar mass as early as $z\sim9$, enabling the identification of galaxies that were missed by the Lyman Break technique used to identify samples with {\it{HST}} \citep{Weibel2024} as well as  the very faintest, low mass galaxies \citep{Navarro-Carrera2024}. Some early JWST results have been unexpected, and are in tension with predictions from models. These include the particularly high stellar masses derived for some high-redshift candidates \citep[e.g.][]{Labbe2023,Rodighiero2023,Xiao2023}, which sparked suggestions of either a crisis with the Lambda Cold Dark Matter ($\Lambda$CDM) cosmological model \citep{Boylan-Kolchin2023,Lovell2023} or evidence of more exotic physical scenarios, such as a variable stellar initial mass function (IMF: \citealt{Steinhardt2022,Trinca2023}). \\
\indent However, deriving stellar masses from photometry is subject to uncertainties in stellar evolution, the stellar IMF, the treatment of interstellar dust, AGN `contamination', and the form of the star formation history (SFH), all of which are known to hamper accurate estimates \citep{Conroy2009,Conroy2010,Walcher2011,Michaowski2012,Michaowski2014,Conroy2013}. One fundamental limitation is the effect of `outshining', whereby younger, brighter stars dominate a galaxy's SED and hide emission from older stars, leading to an {\it{underestimation}} of the total stellar mass \citep[e.g.][]{Sawicki1998,Papovich2001,Pforr2012,Pforr2013,Sorba2015,Sorba2018,Gimenez-Arteaga2023,Gimenez-Arteaga2024}. Another limitation is degenerate physical scenarios leading to similar photometry; for some of the high-redshift galaxies identified by JWST, strong emission lines, dust reddening or an AGN have mimicked the photometric signatures of a genuine evolved stellar population and potentially resulted in significantly {\it{overestimated}} stellar masses \citep{Schaerer2009,Barro2023,Desprez2023,Matthee2023,Trussler2023}. Tests of the methods used to recover galaxy physical properties are essential to understand the observations, refine fitting techniques, and hence maximise the value of JWST data. \\
\indent A limited amount of previous work has compared galaxy parameter estimates obtained using different Spectral Energy Distribution (SED) fitting codes \citep[e.g.][]{Mobasher2015,Hunt2019,Cochrane2021,Pacifici2022,Best2023} and different prior assumptions \citep[e.g.][]{Buat2014}. However, these tend to be tuned to specific populations - e.g. radio-selected galaxies as in \cite{Best2023}. Modelled galaxies - where the `ground truth' physical properties are known and observed-frame photometry or spectra are forward-modelled - offer an alternative test bed for SED fitting. Basic tests have involved generating simplified parametric SFH models, and then refitting these to test the recovery of the input parameters \citep{Bisigello2019,Carnall2019,Suess2022}. Other studies have used mock galaxies drawn from semi-analytic models, with emission modelled using stellar population synthesis alongside simple dust models \citep{Pforr2012,Pforr2013,Mobasher2015}.\\
\indent Hydrodynamical simulations provide more sophisticated models for star formation and feedback, resulting in more complex star formation histories, which may not be well-represented by commonly-used parametric SFHs \citep[e.g.][]{Simha2014}. Depending on resolution, such simulations can also provide information on the dust-star geometry, enabling more detailed treatments of dust attenuation \citep[e.g. see][]{Cochrane2019,Cochrane2023b,Cochrane2023d}. Hydrodynamical simulations, with photometry generated via radiative transfer post-processing, have hence been used to test the fidelity of galaxy parameter recovery with common SED fitting codes \citep[e.g.][]{Wuyts2009,Michaowski2014,Hayward2014,Smith2017,Haskell2023,Haskell2024}. Although larger box cosmological hydrodynamical simulations are typically lower resolution, they naturally provide larger samples, enabling studies of parameter recovery across broader parameter space \citep[e.g.][]{Dudzeviciute2019,Katsianis2020}. \cite{Lower2020} performed such a study using $z\sim0$ modelled galaxies drawn from the SIMBA simulation, demonstrating that inferred stellar masses are sensitive to the SFH parametrisation. They found that the average offset between inferred and true stellar mass was $0.4\,\rm{dex}$ with a delayed-$\tau$ model, but just $-0.02\,\rm{dex}$ with a non-parametric implementation \citep[e.g.][]{Panter2007,Tojeiro2007,Tojeiro2009,Leja2019a}. 
This is consistent with the conclusions drawn by \cite{Leja2019}: stellar mass functions inferred from 3D-{\it{HST}} data using fits with a delayed exponential model are inconsistent with those inferred using a non-parametric model, with the latter resulting in more massive, older galaxies.\\
\indent Given the observational opportunities provided by JWST, in particular its ability to probe rest-frame optical and near-infrared emission for high-$z$ galaxies, it is important to extend these studies and validate the use of SED fitting codes at higher redshifts. Recently, \cite{Narayanan2024} explored this topic using $z=7$ simulated galaxies. Fitting forward-modelled JWST NIRCam and MIRI photometry with the Prospector SED fitting code \citep{Leja2017,Johnson2021}, they found that stellar masses were remarkably poorly recovered, attributing this to outshining. However, outshining could reasonably be expected to  have a smaller effect at high redshift ($z=7$ is only $\sim750\,\rm{Myr}$ after the Big Bang). In their study, stellar masses are recovered to be $10^{9}\,\rm{M_{\odot}}$ regardless of true stellar mass (with these ranging between $10^{8}\,\rm{M_{\odot}}$ and $10^{9}\,\rm{M_{\odot}}$) for several of the assumed SFH models. This has worrying implications for the prospects of measuring stellar masses at early times with JWST. However, these results were not reproduced by \cite{Ciesla2024}, who performed SED fitting on the same simulated galaxies and found that stellar masses were well-recovered, with a systematic offset between true and recovered masses of only $0.07\,\rm{dex}$. Since many studies rely only on NIRCam data, it is also important to test stellar mass recovery without MIRI data. Several studies have argued that the absence of rest-NIR data from MIRI may bias high-$z$ stellar mass estimates high \cite[e.g. see][]{Song2023,Papovich2023}. \\
\indent In this Letter, we explore these issues in more detail using the S{\sc phinx}$^{20}$ cosmological radiation hydrodynamics simulation \citep{Katz2023}. S{\sc phinx}$^{20}$ is an ideal simulation suite for testing SED fitting codes: its high resolution and detailed stellar feedback model makes predictions for both dust-star geometries and fine-grained star formation histories, and its relatively large box size (side length $20\,\rm{comoving\,\,Mpc}$) enables tests over a range of galaxy stellar masses. Our Letter is laid out as follows. In Section \ref{sec:methods}, we describe the methods used to generate and fit synthetic photometry for simulated galaxies drawn from S{\sc phinx}$^{20}$. In Section \ref{sec:results}, we quantify the robustness of the stellar mass recovery for synthetic galaxy samples at different redshifts, for different assumed SFH models. In Section \ref{sec:discussion}, we discuss our results in light of previous work. We draw conclusions in Section \ref{sec:conclusions}.

\section{Methods}\label{sec:methods}
We adopt an approach that is similar to previous work that has tested the fidelity of SED fitting codes on forward-modelled galaxies drawn from cosmological hydrodynamical simulations \citep[e.g.][]{Lower2020,Haskell2023,Haskell2024,Narayanan2024}. In short, we draw simulated galaxies, with forward-modelled spectra generated using radiative transfer, from the S{\sc phinx}$^{20}$ simulation. We convolve these spectra with JWST NIRCam filter transmission curves to obtain synthetic photometry, which we then fit with an SED fitting code given varying assumptions. In this section, we describe each of these steps in detail.

\subsection{S{\sc phinx}$^{20}$ simulation}
We draw simulated galaxies from the S{\sc phinx}$^{20}$ cosmological radiation hydrodynamics simulation (using version 1 of the S{\sc phinx} Public Data Release; \citealt{Katz2023}\footnote{\url{https://github.com/HarleyKatz/SPHINX-20-data}}), which comprises a volume of $(20\,\rm{cMpc})^{3}$. S{\sc phinx}$^{20}$ is the largest volume simulation in the S{\sc phinx} suite (see also \citealt{Rosdahl2018,Rosdahl2022}), and also currently the largest cosmological radiation hydrodynamics simulation that resolves a multi-phase ISM. It hence provides a sizeable sample of well-resolved galaxies for our study. We briefly reproduce the key features of the simulation here, but refer the reader to \cite{Katz2023} for more details.\\
\indent S{\sc phinx}$^{20}$ was run using the R{\sc amses} adaptive mesh refinement code \citep{Teyssier2002}, using cosmological initial conditions selected to produce a typical patch of the Universe (i.e. not an especially over- or under-dense region). The dark matter particle mass is $2\times10^{5}\,\rm{M_{\odot}}$, and the initial stellar particle mass is $400\,\rm{M_{\odot}}$. Star formation is modelled using a variable local efficiency, dependent on the thermo-turbulent properties of the gas, with supernova feedback modelled following \cite{Kimm2014}. Lyman continuum radiation from stellar particles is modelled using BPASS v2.2.1 \citep{Stanway2018}. A radiation-hydrodynamics scheme models the propagation of Lyman continuum photons \citep{Rosdahl2015,Rosdahl2018}. S{\sc phinx}$^{20}$ does not include prescriptions for black hole formation, growth or feedback. Our sample comprises all of the S{\sc phinx}$^{20}$ galaxies at $z\geq5$ for which spectra are publicly available: there are $49, 66, 128, 177, 276$ and $317$ galaxies at $z=10, 9, 8, 7, 6$ and $5$, respectively. These were selected by \cite{Katz2023} as those with $10\,\rm{Myr}$-averaged SFR $\geq0.3\,\rm{M_{\odot}yr^{-1}}$, a criterion that roughly selects populations that are observable with JWST. 

\subsection{Radiative transfer post-processing and synthetic JWST photometry}\label{sec:photometry}
We use the spectra released by \cite{Katz2023} as our starting point for generating photometry in JWST's NIRCam bands. We provide a brief summary of the methods used to generate these spectra here.
Intrinsic stellar continuum emission was generated based on particle age, mass and metallicity and using the BPASS v2.2.1 \citep{Stanway2018} models. The generation of emission line luminosities follows the methods described in \cite{Choustikov2024}, which are based on {\small{CLOUDY}} models \citep{Ferland2017}. Nebular continuum emission is also modelled (with free-free and two-photon emission calculated following \cite{Schirmer2016} and free-bound emission calculated with CHIANTI, as described by \citealt{Dere2019}). Dust-attenuated spectra are then modelled using the Monte-
Carlo radiative transfer code R{\sc ascas} \citep{Michel-Dansac2020}, assuming an SMC dust model \citep{Gordon2003}, using the methods described by \cite{Laursen2009}. \\
\indent Observed-frame, dust-attenuated spectra were generated along 10 fixed lines of sight and made publicly available as part of the S{\sc phinx}$^{20}$ data release. In this work, we select a random line of sight for each galaxy and generate JWST photometry from the appropriate dust-attenuated spectrum using the {\it{pyphot}} python package. We match the NIRCam photometric coverage of the Public Release IMaging for Extragalactic Research (PRIMER; GO 1837; PI Dunlop), a major public JWST Treasury Program. PRIMER is imaging $378\,\rm{arcmin}^{2}$ within UDS and COSMOS with eight NIRCam filters (F090W, F115W, F150W, F200W, F277W, F356W, F410M and F444W), and $237\,\rm{arcmin}^{2}$  with two MIRI filters (F770W and F1800W). Since S{\sc phinx}$^{20}$ spectra are generated only out to a rest-frame wavelength of $\sim1\,\mu\rm{m}$, we do not include synthetic MIRI data in this study. Following \cite{Narayanan2024}, we consider the `ideal case' to characterise the best possible performance of the SED fitting models. Hence, we do not add noise to the photometry.

\begin{figure*}
\begin{subfigure}[t]{0.32\textwidth}
\includegraphics[width=1\columnwidth]{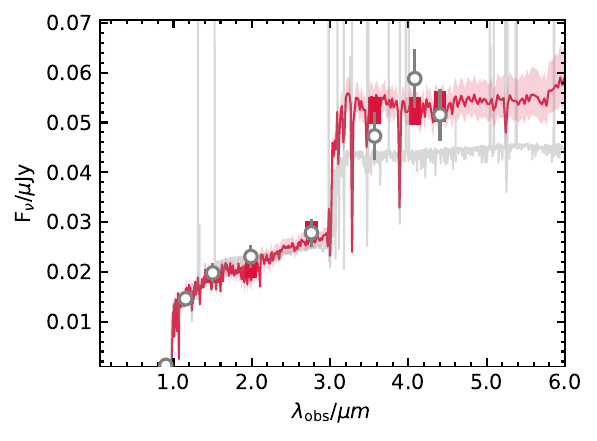}
\includegraphics[width=1\columnwidth]{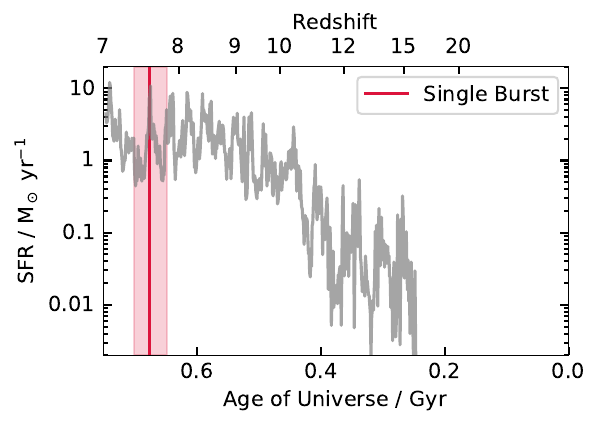}
\caption{Single Burst SFH} \label{fig:panela}
\end{subfigure}
\begin{subfigure}[t]{0.32\textwidth}
\includegraphics[width=1\columnwidth]{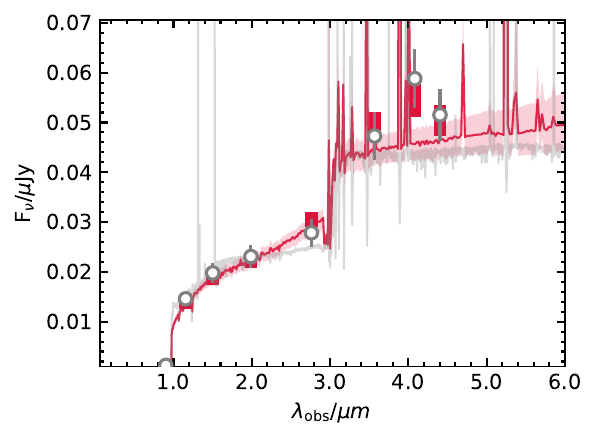}
\includegraphics[width=1\columnwidth]{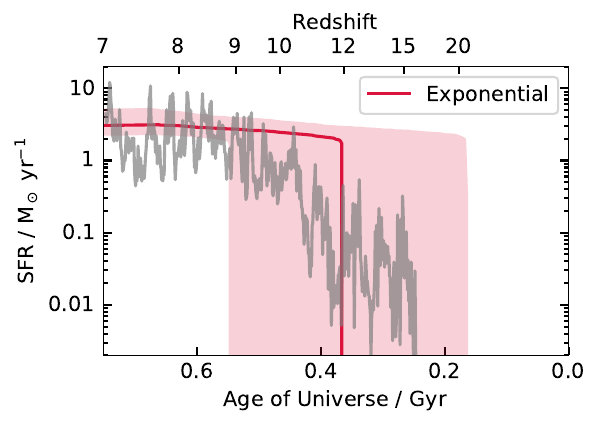}
\caption{Exponential SFH} \label{fig:panelb}
\end{subfigure}
\begin{subfigure}[t]{0.32\textwidth}
\includegraphics[width=1\columnwidth]{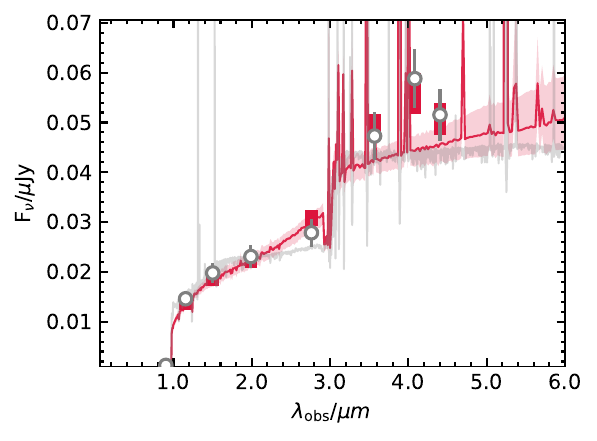}
\includegraphics[width=1\columnwidth]{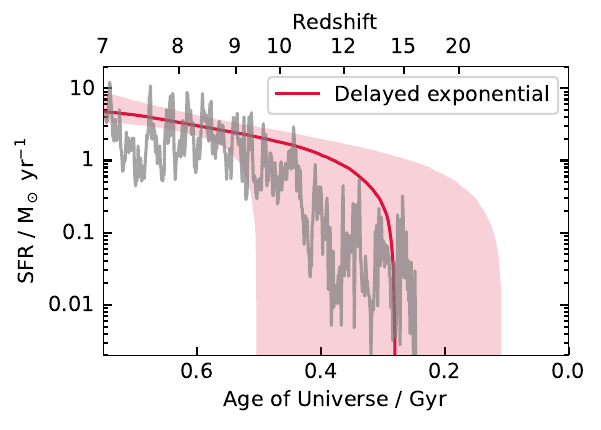}
\caption{Delayed Exponential SFH} \label{fig:panelc}
\end{subfigure}

\vspace{0.5cm}

\begin{subfigure}[t]{0.32\textwidth}
\includegraphics[width=1\columnwidth]{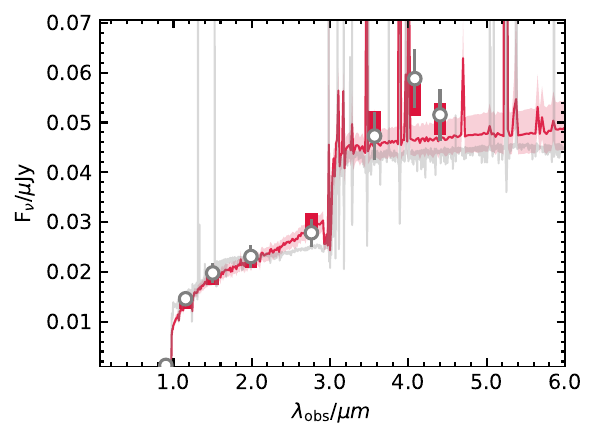}
\includegraphics[width=1\columnwidth]{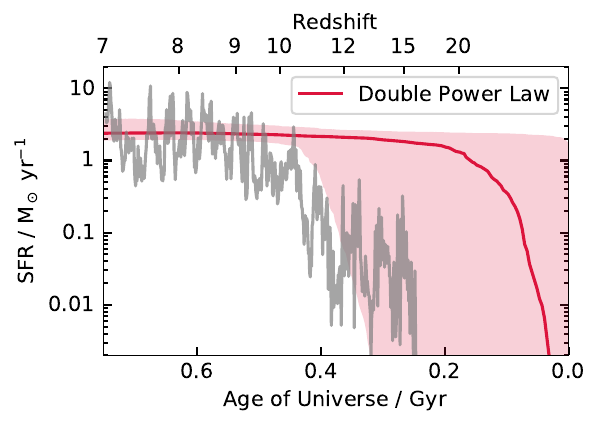}
\caption{Double Power Law SFH} \label{fig:paneld}
\end{subfigure}
\begin{subfigure}[t]{0.32\textwidth}
\includegraphics[width=1\columnwidth]{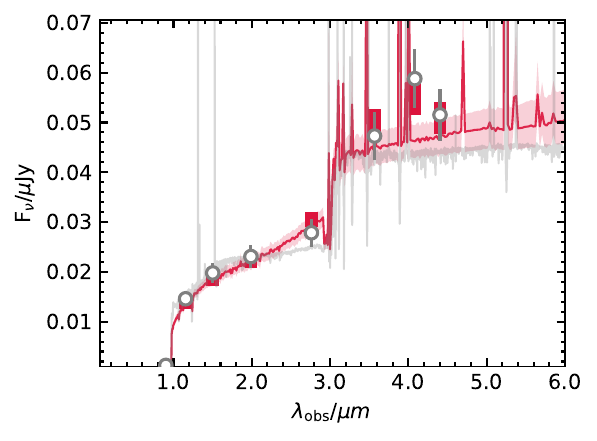}
\includegraphics[width=1\columnwidth]{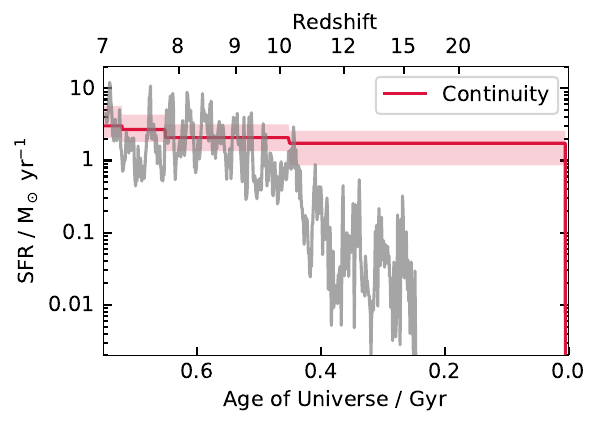}
\caption{Continuity SFH} \label{fig:panele}
\end{subfigure}
\begin{subfigure}[t]{0.32\textwidth}
\includegraphics[width=1\columnwidth]{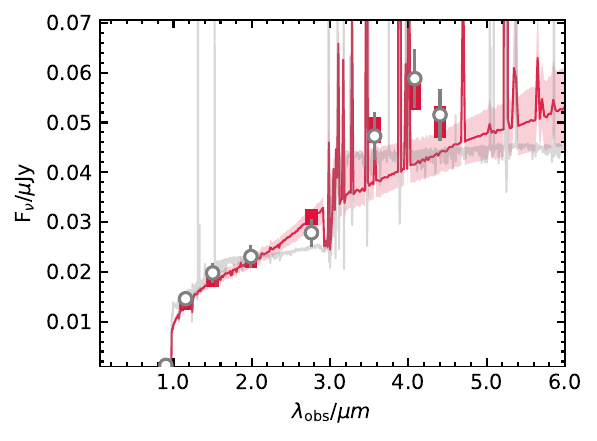}
\includegraphics[width=1\columnwidth]{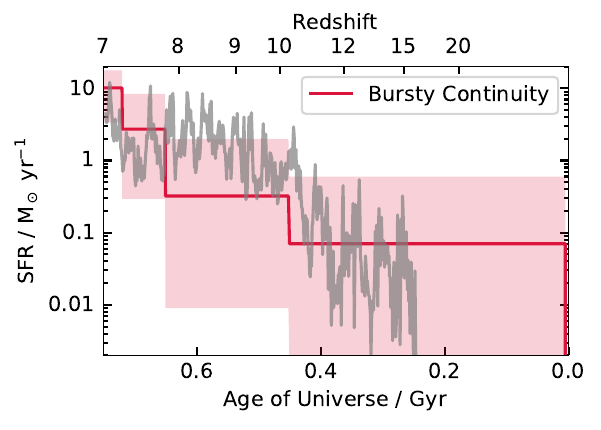}
\caption{Bursty Continuity SFH} \label{fig:panelf}
\end{subfigure}
\caption{Example of fits (pink shaded regions) to the modelled NIRCam photometry (grey points with error bars) of a $z=7$ simulated galaxy (full spectrum shown in grey) along a single line of sight (upper panels), and posterior star formation histories (lower panels), given different assumed SFH parametrisations (see labels). The data are similarly well-modelled by photometry generated using very different SFH parametrisations.}
\label{fig:fitted_sed_example_135}
\end{figure*}

\subsection{Spectral Energy Distribution fitting}\label{sec:sed_fitting}
We use the SED fitting code Bayesian Analysis of Galaxies for Physical Inference and Parameter EStimation ({\small{BAGPIPES}}; \citealt{Carnall2018}) to fit the forward-modelled NIRCam photometry using multiple star formation history parametrisations. Here, we describe our methods, first outlining the common parameter choices that remain fixed between these SFH iterations, and then detailing the SFH parametrisations themselves.\\
\indent Our fitting uses the fiducial simple stellar population (SSP) models implemented by {\small{BAGPIPES}}, the 2016 version of the BC03 templates \citep{Charlot2003}, with a \cite{Kroupa2002} IMF. Note that these are different templates to those used in the simulations (as in observational studies, where the `true' SSPs are not known). Nebular emission is computed using the {\small{CLOUDY}} photoionization code \citep{Ferland2017}, following \cite{Byler2017}. {\small{CLOUDY}} is run using each SSP template as the input spectrum. Dust grains are included using {\small{CLOUDY}}'s `ISM' prescription, which implements a grain-size distribution and abundance pattern that reproduces the observed extinction properties for the ISM of the Milky Way. We fit the slope of the dust attenuation curve using a \cite{Salim2018} parametrisation (see \citealt{Meldorf2024} and \citealt{Osborne2024} for detailed studies of parameter recovery using this flexible dust model). This enables the deviation of the slope from the \cite{Calzetti2000} dust attenuation curve to be fitted. This is particularly important for fitting low-mass galaxies with low dust attenuation, which typically have steeper dust attenuation laws \citep{Witt2000}. We fit the normalisation of the curve, the $V-$band dust attenuation, using the prior $A_{V}=[0.0,2.0]$, and fix the $2175\,\angstrom$ bump strength to zero. We assume that birth clouds experience additional dust attenuation, so that young stars are attenuated twice as much as old stars \citep{Calzetti1994}.\\
\indent For all models, we set a uniform prior of $[5.0,12.0]$ for the logarithm of the total stellar mass formed in solar masses. We set the metallicity prior to be uniform in log space, $Z=[0.0001,2.0]\,Z_{\odot,\rm{old}}$, where $Z_{\odot,\rm{old}}$ denotes solar models prior to \cite{Asplund2009}. We fix the redshift at the true value, following the `best-case scenario' approach of \cite{Narayanan2024}.\\
\indent We test several star formation history models, four parametric (simple burst, exponential, delayed-tau, and a more flexible double power law), and two non-parametric (the `continuity' model and the `bursty continuity' model). These models are defined as follows:\\

\begin{table*}[]
    \centering
    \begin{tabular}{l|c|c|c|c|c|c|c}
         SFH model & $z=5$ & $z=6$ & $z=7$ & $z=8$ & $z=9$ & $z=10$ \\
         \hline
         (a) Single Burst & [$-0.02$, $0.23$] & [$-0.05$, $0.15$] & [$0.13$, $0.21$] & [$0.10$, $0.19$] & [$0.11$, $0.16$] & [$0.19$, $0.15$]\\
         (b) Exponential & [$0.04$, $0.24$] & [$0.07$, $0.14$] & [$0.17$, $0.20$] & [$0.33$, $0.21$] & [$0.25$, $0.15$] & [$0.37$, $0.14$]\\
         (c) Delayed Exponential & [$-0.07$, $0.25$] & [$0.00$, $0.16$] & [$0.10$, $0.20$] & [$0.32$, $0.20$] & [$0.22$, $0.14$] & [$0.28$, $0.15$]\\
         (d) Double Power Law & [$0.09$, $0.24$] & [$0.11$, $0.15$] & [$0.20$, $0.19$] & [$0.32$, $0.21$] & [$0.27$, $0.14$] & [$0.41$, $0.15$] \\
         (e) Continuity & [$0.13$, $0.21$] & [$0.12$, $0.14$] & [$0.19$, $0.20$] & [$0.35$, $0.21$] & [$0.30$, $0.15$] & [$0.41$, $0.15$]\\
         (f) Bursty Continuity & [$0.07$, $0.17$] & [$0.00$, $0.14$] & [$-0.01$, $0.16$] & [$0.27$, $0.19$] & [$0.17$, $0.12$] & [$0.15$, $0.13$]\\
    \end{tabular}
    \caption{Characterisation of stellar mass recovery for the whole sample at each redshift, when fitted with the six SFH models. We present [median ($\Delta M_{\star}$), standard deviation ($\Delta M_{\star}$)]. Positive $\Delta M_{\star}$ values correspond to overestimated stellar masses. Stellar masses are generally well-recovered when averaged across the population, with a median systematic offset less than $0.3\,\rm{dex}$ (a factor of two) in most cases, and similar scatter. However, there are significant trends within the population that are not illuminated by these population-averaged measures (see Section \ref{sec:fitted_mass_trends}). \vspace{0.2cm}}
    \label{tab:mass_fits}
\end{table*}

\noindent{\bf{a) Single burst}}\\
In this parametrisation, all the stellar mass is formed in a single delta function burst; the age of this burst is the only fitted SFH parameter.\\\\
{\bf{b) Exponential}}\\
Here, the star formation history is parametrised using:
\begin{equation}
{\rm{SFR}}(t) \propto e^{-t/\tau},
\end{equation}
where $\tau$ is the characteristic $e$-folding time with which the SFR declines. This parametrisation has two free parameters: $\tau$ and the age of the galaxy. \\\\
{\bf{c) Delayed Exponential}}\\
Here, the star formation history is parametrised using:
\begin{equation}
{\rm{SFR}}(t) \propto t e^{-t/\tau},
\end{equation}
where $\tau$ is, again, the characteristic $e$-folding time. Note that both Exponential and Delayed Exponential parametrisations are forms of the Gamma Distribution. \\\\
{\bf{d) Double Power Law}}\\
Here, the star formation history is parametrised using:
\begin{equation}
    {\rm{SFR}}(t) \propto \Bigg[\Big(\frac{t}{\tau}\Big)^{\alpha} + \Big(\frac{t}{\tau}\Big)^{-\beta} \Bigg]^{-1},
\end{equation}
where $\alpha$ is the slope in the region of falling SFR, and $\beta$ is the slope in the region of rising SFR. $\tau$ relates to the time at which the SFR peaks. The priors for both $\alpha$ and $\beta$ are set to be uniform in log space, in the range $[0.01, 1000]$.\\\\
{\bf{e) Continuity}}\\
Our final two models are `non-parametric', i.e. rather than assuming a functional form for the SFH, they fit star formation within time bins. The continuity prior, introduced by \cite{Leja2019}, fits $\Delta\log(\rm{SFR})$ between adjacent time bins, weighting against sharp changes in ${\rm{SFR}}(t)$. The prior on $x=\log({\rm{SFR}}_{n}/{\rm{SFR}}_{n+1})$ is described by the Student's $t$-distribution:\\
\begin{equation}
{\rm{PDF}}\left(x,\nu\right)=\frac{\Gamma\left(\frac{\nu+1}{2}\right)}{\sqrt{\nu\pi}\,\Gamma\left(\frac{\nu}{2}\right)}\left(1+\frac{\left(x/\sigma\right)^{2}}{\nu}\right)^{-\frac{\nu+1}{2}},
\end{equation}
where $\Gamma$ is the Gamma function, $\sigma$ is a scale factor controlling the width of the distribution, and $\nu$ is the number of degrees of freedom controlling the probability in the tails of the distribution. The Student's $t$-distribution has heavier tails than the normal distribution. Parameters are set to $\sigma=0.3$ and $\nu=2$, following \cite{Leja2019}\footnote{Note that there is a typo in the value \cite{Narayanan2024} provide for $\sigma$ ($3$ rather than $0.3$), but the correct value was used in their fitting.}. These parameters are motivated by the star formation histories of simulated galaxies \citep{Leja2019a}. We set time bins to: $[0,10,30,100,300,t_{\rm{max}}]\,\rm{Myr}$, where $t_{\rm{max}}$ is the age of the Universe at the redshift studied.\\\\
{\bf{f) Bursty Continuity}}\\
Here, the parameters in the Student's t-distribution are adjusted to $\sigma=1$ and $\nu=2$, following \cite{Tacchella2022}. This relaxes the continuity condition, enabling more dramatic changes in star formation rate between time bins, as is predicted by higher-resolution simulations with detailed stellar feedback models \cite[e.g.][]{Sun2023}. We use the same time bins as for the standard Continuity model.\\
\indent We fit the modelled NIRCam photometry (described in Section \ref{sec:photometry}) for each simulated galaxy with all six of these models. In each case, we use the noise-free photometry, with the uncertainty on each measurement set to $10\%$, and the redshift fixed at the true redshift (i.e. $z=10$, $z=9$, $z=8$, $z=7$, $z=6$, or $z=5$). This follows the method of \cite{Narayanan2024}, and enables us to isolate the impact of choice of star formation history on galaxy parameter recovery in the `best case scenario' of perfect photometry. A single SED fit for a modelled galaxy took a few minutes on a laptop. Examples of fitted models with input photometry overlaid are shown in Figure \ref{fig:fitted_sed_example_135}; generally, fits are able to reproduce the input photometry to within the uncertainty supplied. However the set of photometry tested here is not sufficient to discriminate between models; as seen in Figure \ref{fig:fitted_sed_example_135}, very different star formation history models fit the photometry approximately equally well. This has been noted in previous work \citep[e.g.][]{Tacchella2022}. To assess the fits quantitatively, we adopt the Bayesian Information Criterion (BIC; \citealt{Schwarz1978}) test to assess the success of the different SFHs. The BIC is defined as ${\rm{BIC}} = k\log{n}-2\Lagr$, where $k$ is the number of degrees of freedom in the model, $n$ is the number of photometric data points (which is the same for all models), and $\Lagr$ is the log-likelihood. At each redshift, we calculate the BIC for every fitted galaxy, for all six model fits. We then compare the distributions of BIC values obtained for the six SFH parametrisations. Interestingly, the BIC actually favours the simplest model, the single burst SFH parametrisation. This is because all SFH parametrisations are able to fit the data similarly well, so the BIC is driven by the number of fitted parameters (which is lowest for the single burst parametrisation). We quantify the success of each model in recovering the `true' stellar masses and star formation rates of the modelled galaxies in Section \ref{sec:results}.

\begin{figure*} 
\begin{subfigure}[t]{1.0\textwidth}
\centering
\includegraphics[width=0.47\columnwidth]{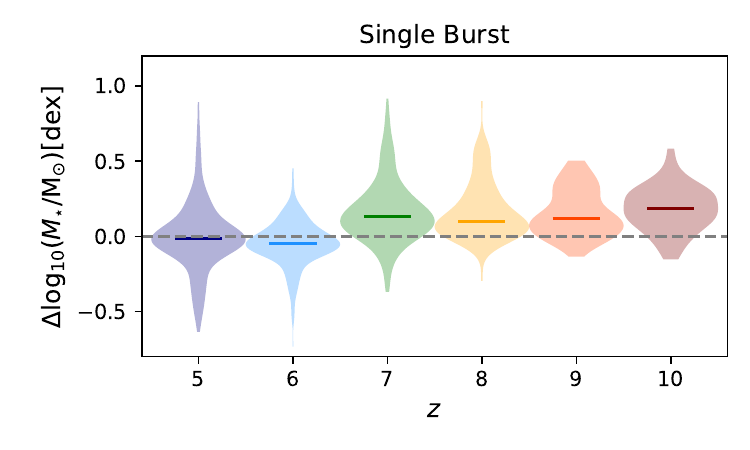}
\includegraphics[width=0.47\columnwidth]{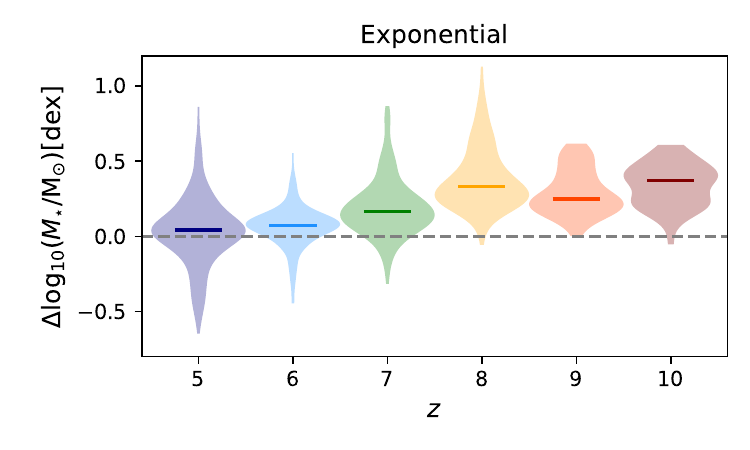}
\includegraphics[width=0.47\columnwidth]{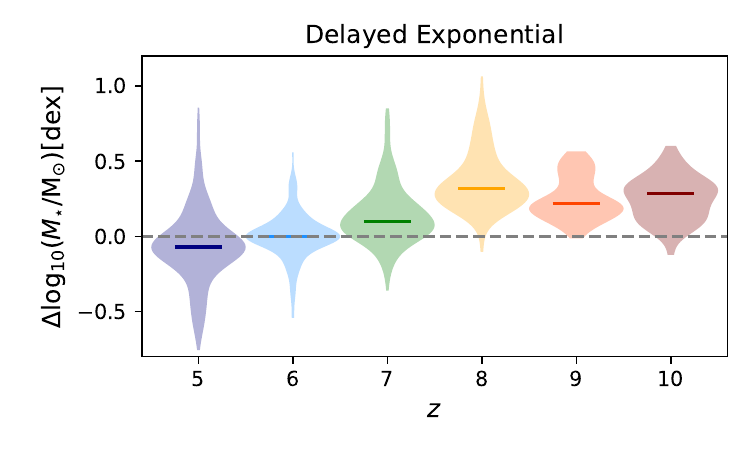}
\includegraphics[width=0.47\columnwidth]{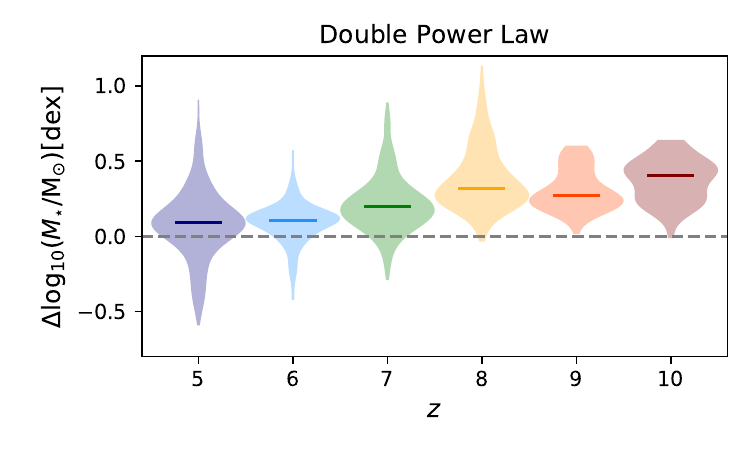}
\includegraphics[width=0.47\columnwidth]{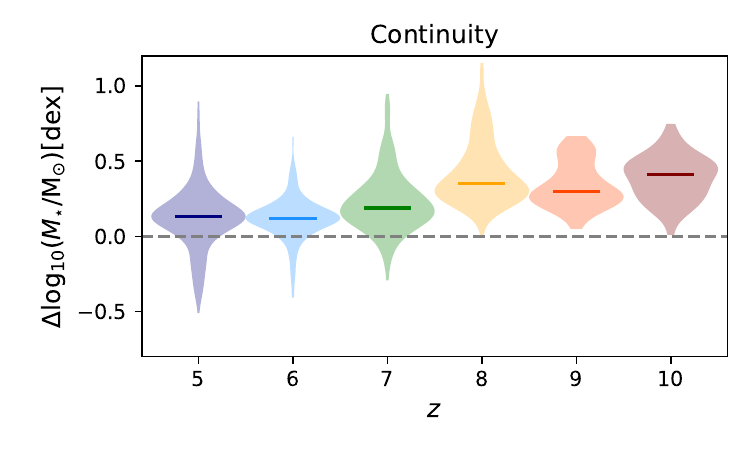}
\includegraphics[width=0.47\columnwidth]{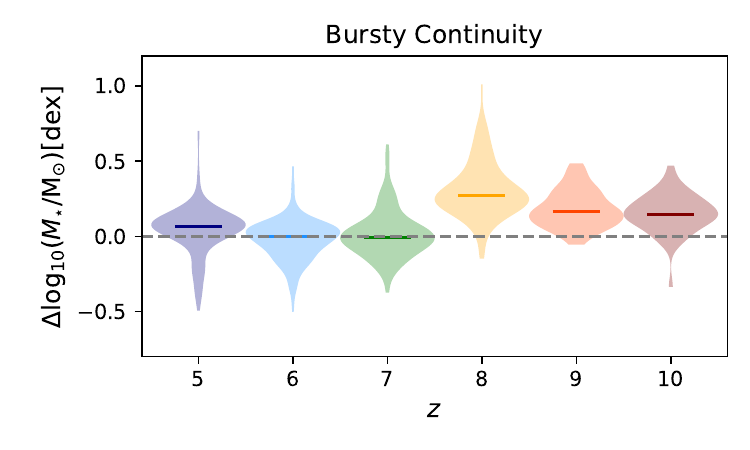}
\end{subfigure}
\vspace{-0.2cm}
\caption{Stellar mass recovery for the whole sample of SPHINX galaxies with $\rm{SFR}>0.3\,\rm{M_{\odot}}yr^{-1}$ at each redshift, when fitted with the six SFH models. The majority of stellar masses are slightly overestimated, but masses are generally recovered to within $\sim0.5\,\rm{dex}$, regardless of SFH parametrisation. Summary statistics are presented in Table \ref{tab:mass_fits}. \vspace{0.2cm}}
\label{fig:violin_plot}
\end{figure*}

\begin{figure*} 
\includegraphics[width=0.33\columnwidth]{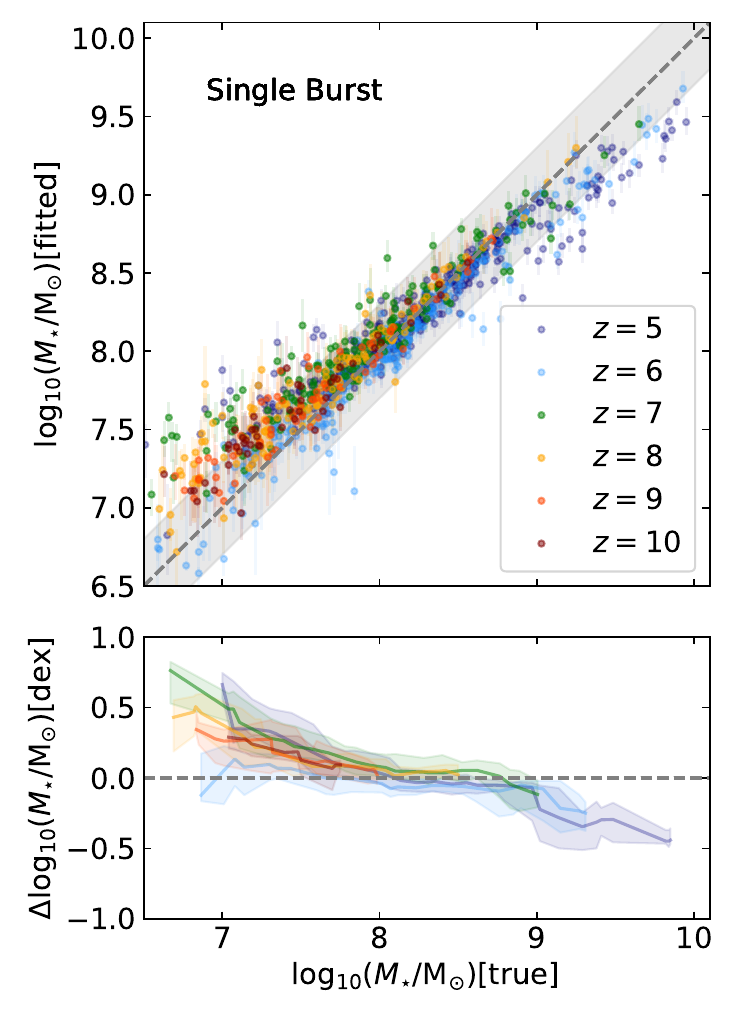}
\includegraphics[width=0.33\columnwidth]{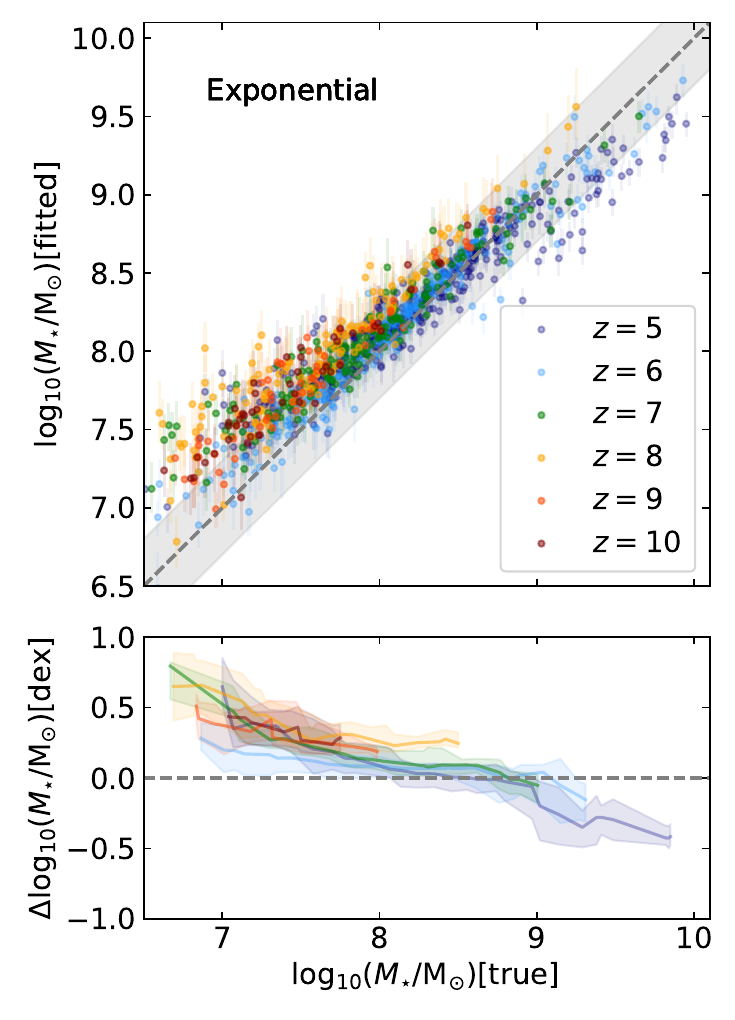}
\includegraphics[width=0.33\columnwidth]{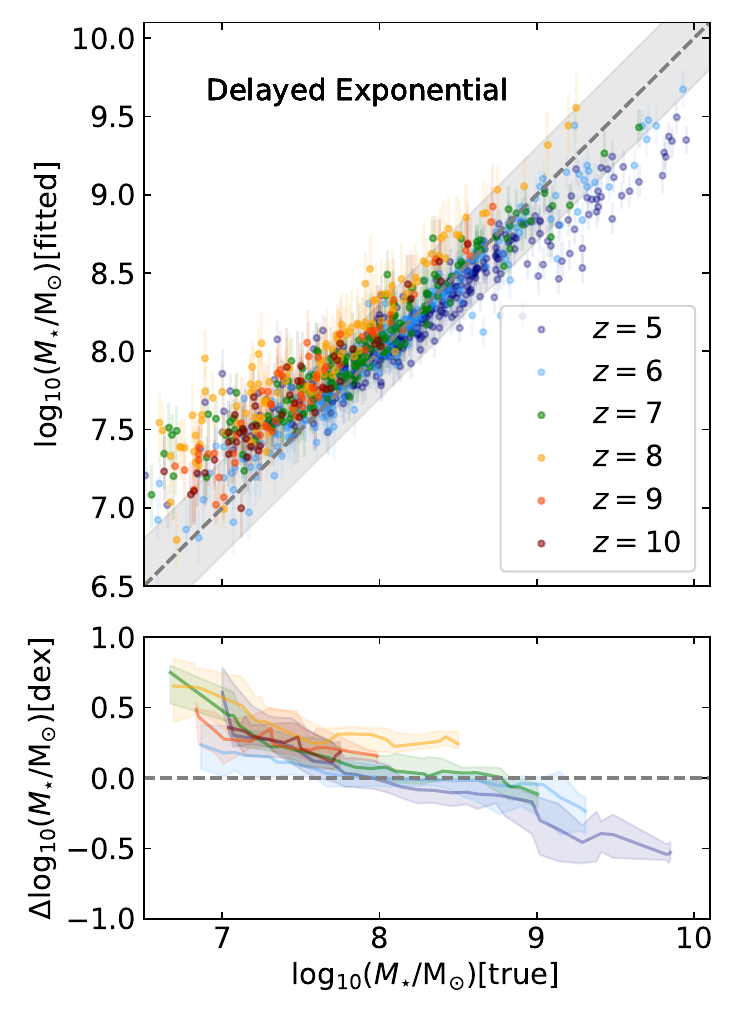}
\includegraphics[width=0.33\columnwidth]{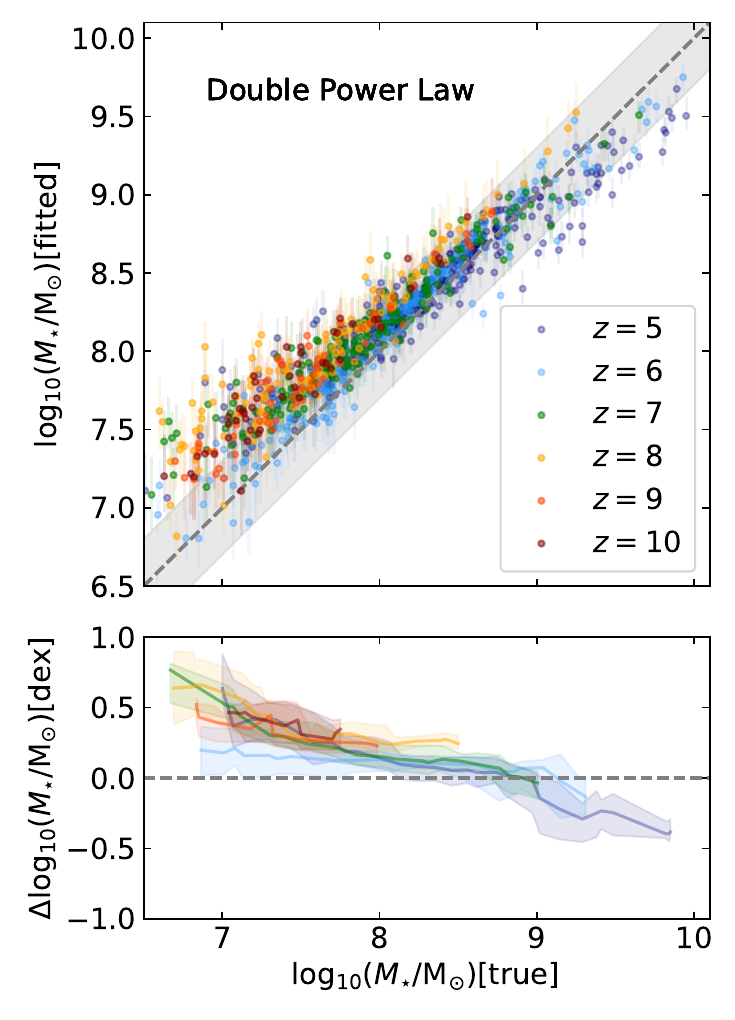}
\includegraphics[width=0.33\columnwidth]{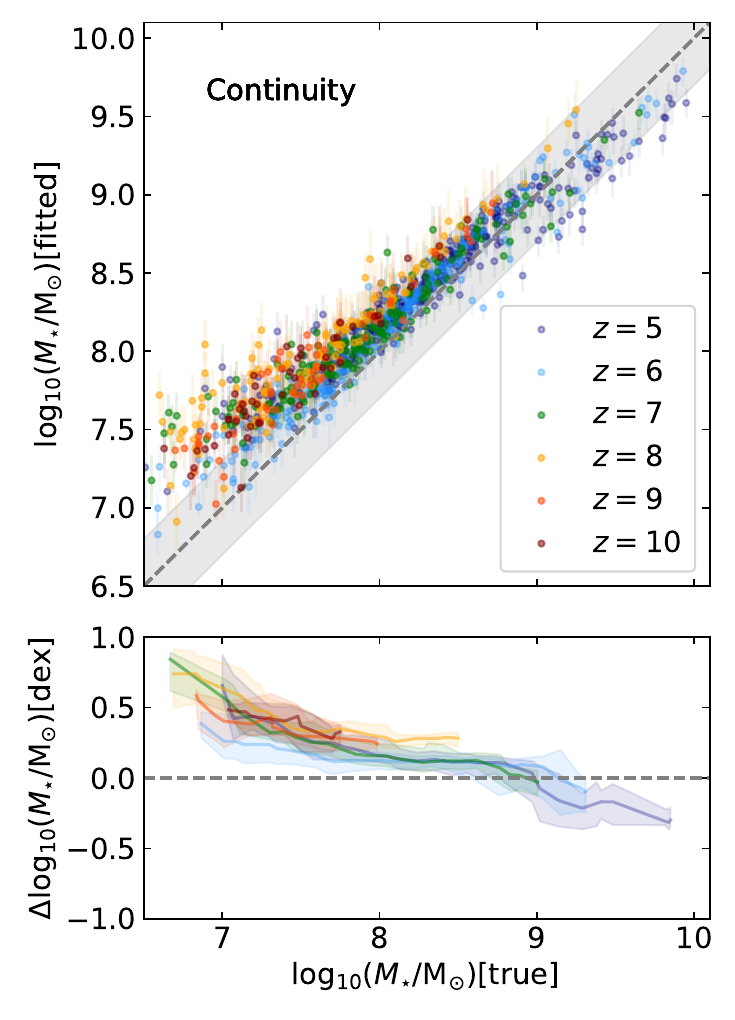}
\includegraphics[width=0.33\columnwidth]{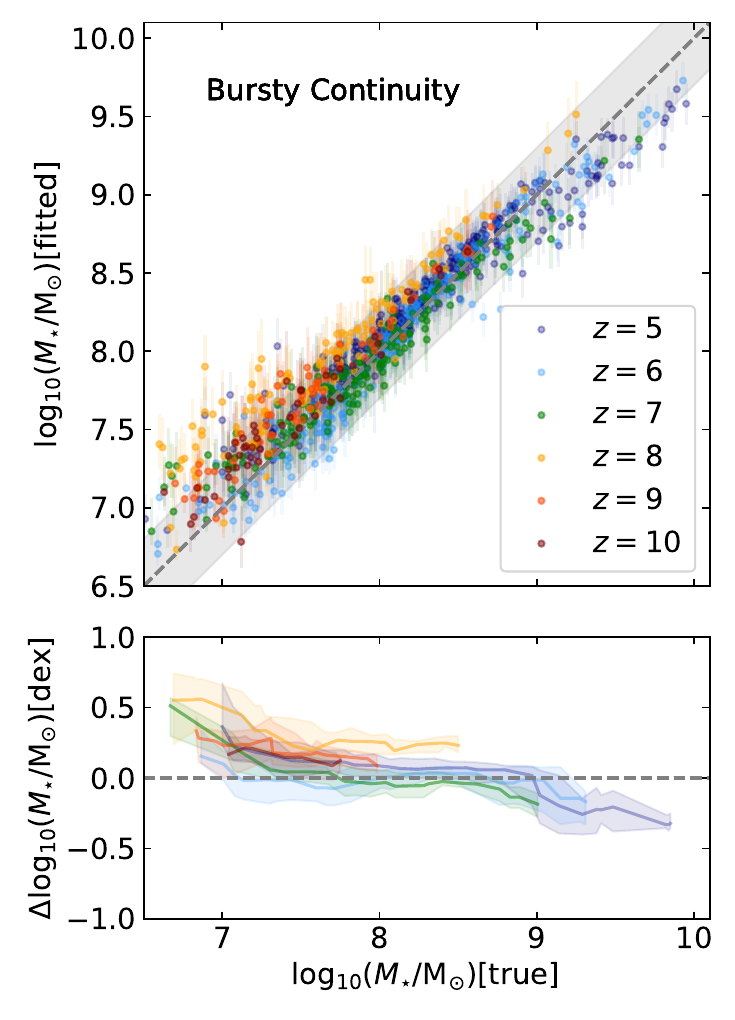}
\caption{Fitted stellar mass (upper panels) and offset between fitted and true stellar mass (lower panels), versus true stellar mass, for the different SFH models (labelled) and samples at different redshifts (see colored points). The shaded grey regions mark out a factor of 2 ($0.3\,\rm{dex}$) above and below the true stellar mass, to guide the eye. The robustness of stellar mass recovery is a function of mass: stellar mass is overestimated at low stellar masses ($M_{\star}\lesssim10^{8}\,\rm{M_{\odot}}$) and underestimated at high stellar masses ($M_{\star}\gtrsim10^{9}\,\rm{M_{\odot}}$). \vspace{0.2cm}}
\label{fig:mstar}
\end{figure*}

\begin{figure*} 
\includegraphics[width=0.34\columnwidth]{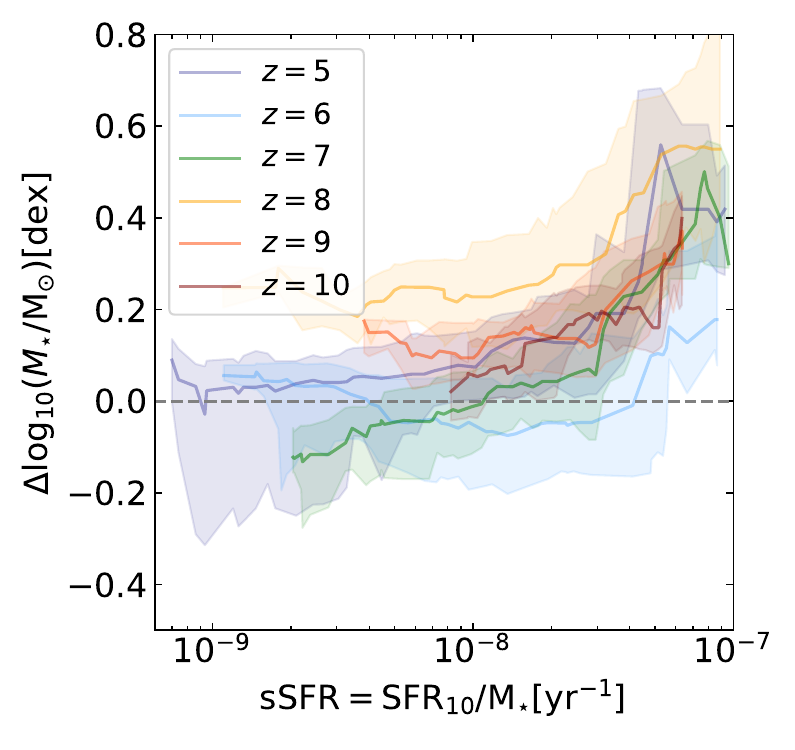}
\includegraphics[width=0.34\columnwidth]{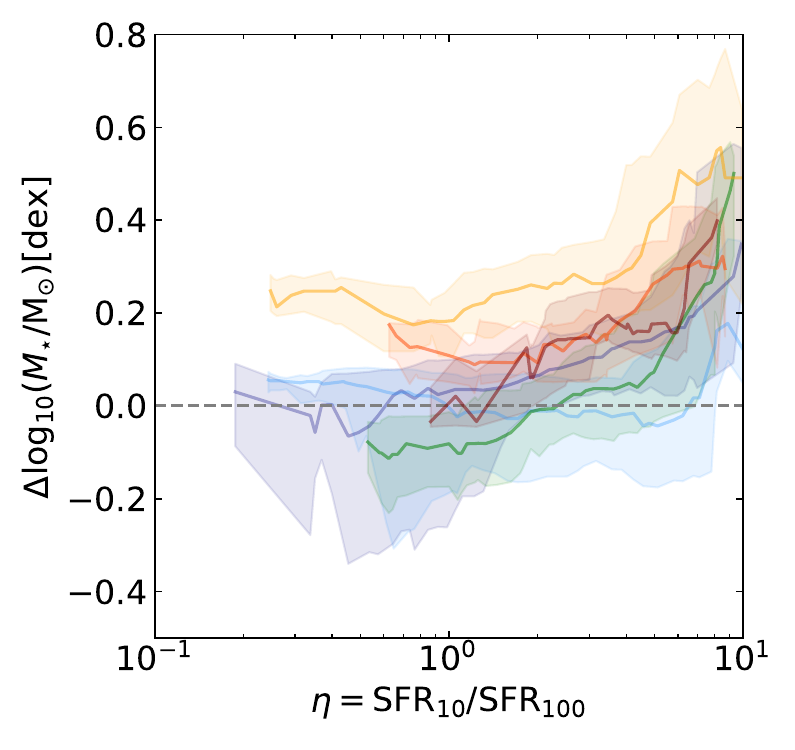}
\includegraphics[width=0.34\columnwidth]{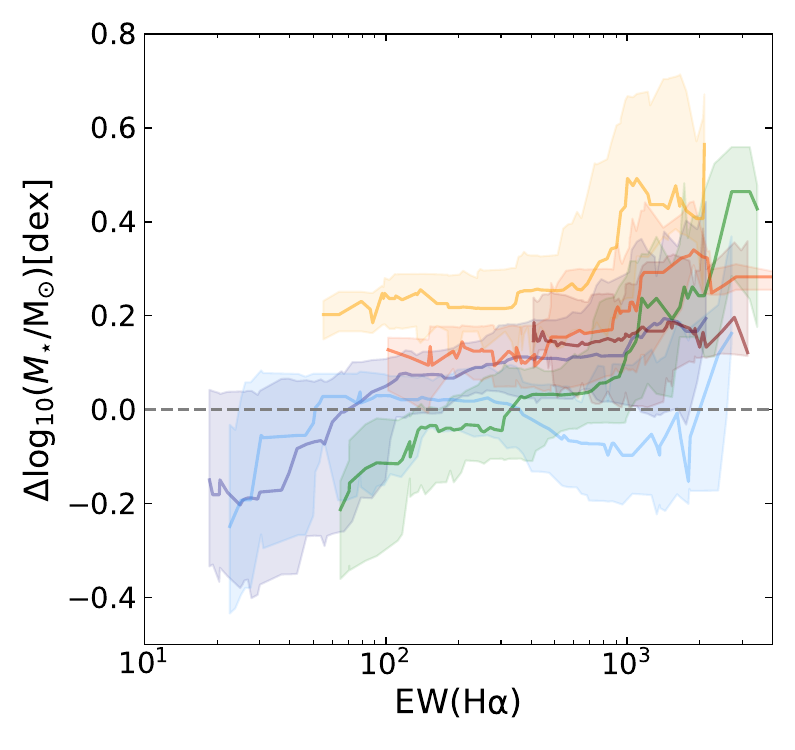}
\includegraphics[width=0.34\columnwidth]{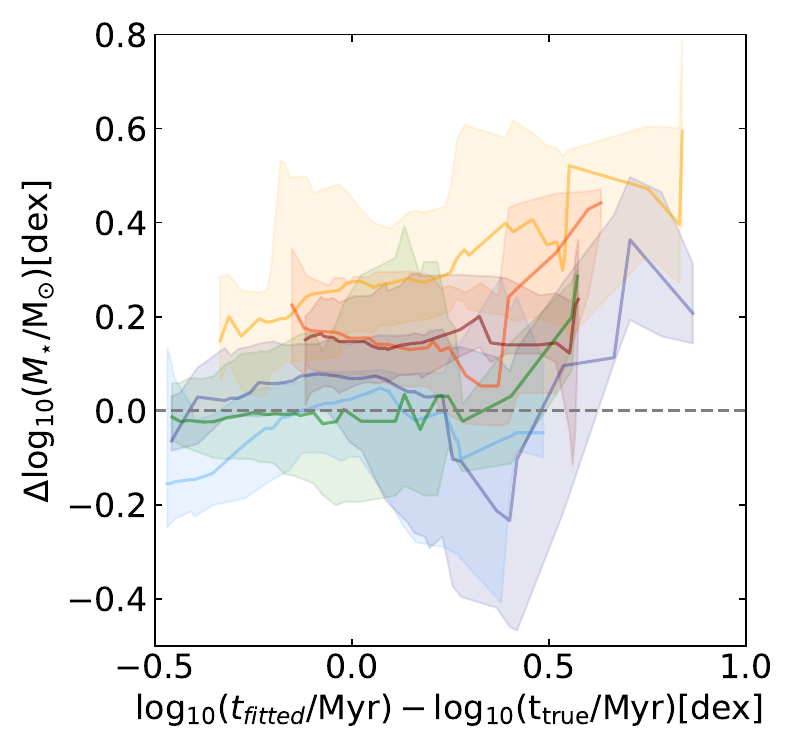}
\caption{Offset between fitted stellar mass and true stellar mass $(\Delta M_{\star} = \log_{10}(M_{\star,\rm{fitted}}/{\rm{M_{\odot}}})-\log_{10}(M_{\star,{\rm{true}}}/\rm{M_{\odot}}))$ as a function of sSFR (upper left), `burstiness' parameter, $\rm{SFR}_{10\rm{Myr}}/\rm{SFR}_{100\rm{Myr}}$ (upper right), $\rm{H}\alpha$ equivalent width (lower left), and offset between fitted and true mass-weighted age (lower right). Each panel shows different redshifts (see colors); all are derived for fits using the Bursty Continuity SFH model. Stellar masses tend to be overestimated where the galaxy has a rising recent SFH ($\rm{SFR}_{10\rm{Myr}}>\rm{SFR}_{100\rm{Myr}}$). Galaxies with overestimated stellar masses also typically have overestimated mass-weighted stellar ages.\vspace{0.2cm}} \label{fig:mstar_vs_properties}
\end{figure*}

\begin{figure*} 
\includegraphics[width=1.0\columnwidth]{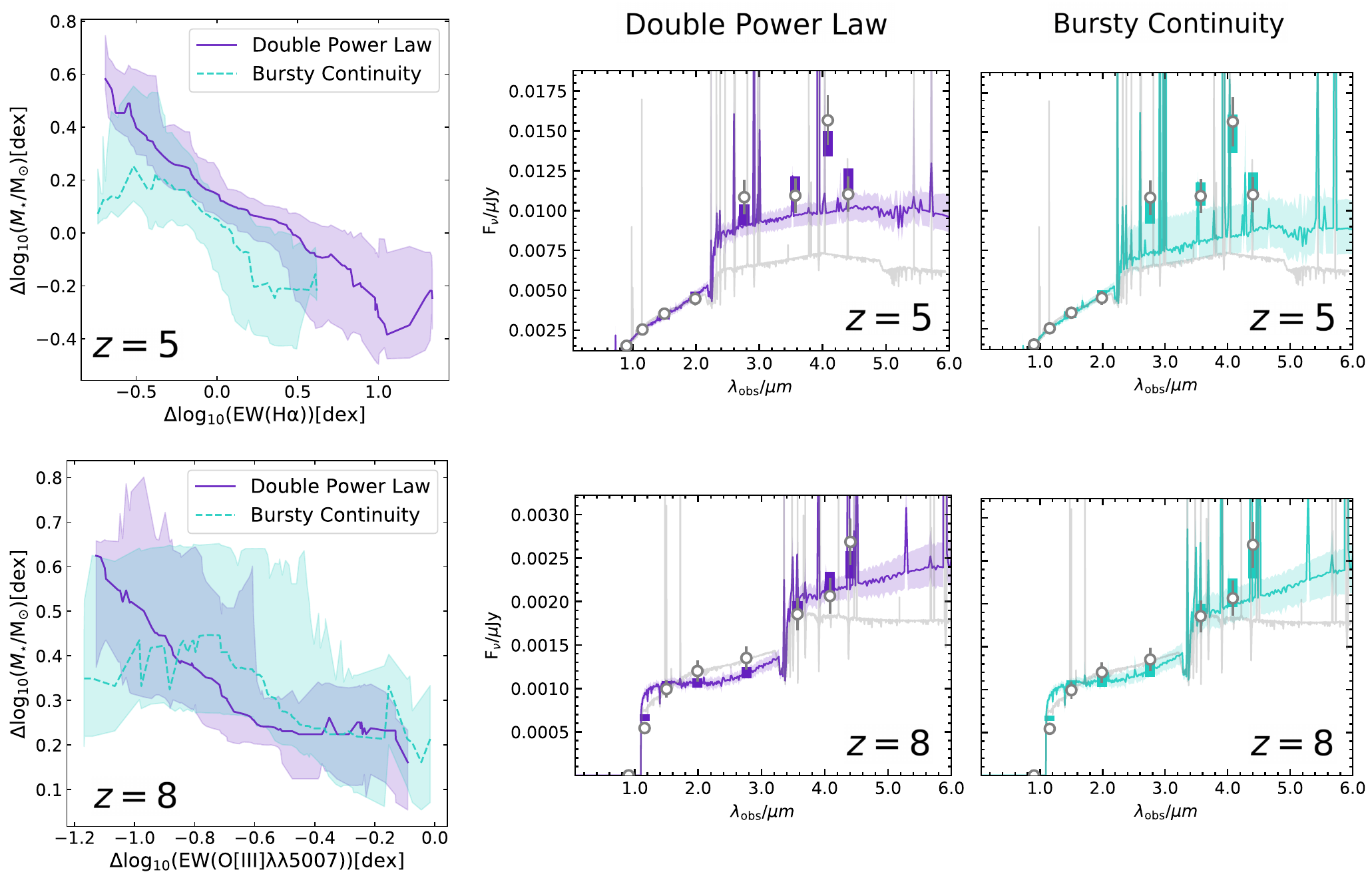}
\vspace{-0.2cm}
\caption{Left panels: offset between fitted stellar mass and true stellar mass versus offset between fitted and true $\rm{H}\alpha$ equivalent width at $z=5$ (upper left panel) and between fitted and true [O{\sc{iii}}]$\lambda5007$ equivalent width, at $z=8$ (lower left panel). Here, we compare fits using the Double Power Law and Bursty Continuity SFH parametrisations. For both parametrisations, when stellar masses are overestimated, emission line equivalent width is underestimated. However, offsets tend to be less severe for fits using the Bursty Continuity parametrisation. Upper central and right-hand panels: an example of a $z=5$ galaxy where the fits (purple and cyan) can diverge significantly from the underlying simulated SED (grey). Strong emission lines ($\rm{H}\alpha$ in F356W, F410M, and F444W and [O{\sc{iii}}]$\lambda4959,5007$ in F277W) are not well-fitted by {\sc{bagpipes}} with the Double Power Law SFH parametrisation; instead, the SED fit favours an older, more massive stellar population with stronger optical continuum and weaker emission lines. As a result, the inferred stellar mass ($\log_{10}(M_{\star}/\rm{M_{\odot}})=8.17^{+0.07}_{-0.12}$) exceeds the true stellar mass ($\log_{10}(M_{\star}/\rm{M_{\odot}})=7.82$). This effect is less strong when the Bursty Continuity SFH is adopted; in this case, the stellar mass is approximately recovered: $\log_{10}(M_{\star}/\rm{M_{\odot}})=7.8\pm0.3$. Lower central and right-hand panels: an example of a $z=8$ galaxy where the fits (purple and cyan) diverge significantly from the underlying simulated SED (grey). The emission lines ([O{\sc{iii}}]$\lambda4959,5007$ and $\rm{H}\beta$ in F444W and [O{\sc{iii}}]$\lambda4363$ in F356W, F410M and F444W) are not well-fitted by {\small{BAGPIPES}} and the SED fit favours an older stellar population. Again, the inferred stellar mass ($\log_{10}(M_{\star}/\rm{M_{\odot}})=7.9\pm0.2$ for both SFH parametrisations) exceeds the true stellar mass ($\log_{10}(M_{\star}/\rm{M_{\odot}})=7.42$).\vspace{0.2cm}}
\label{fig:EL_fits}
\end{figure*}

\section{Results}\label{sec:results}
In this Section, we compare the output of our SED fits against the intrinsic physical properties of the S{\sc phinx}$^{20}$ galaxies. We first present summary statistics for the whole sample at each redshift, and then investigate trends in parameter recovery.
\subsection{The robustness of stellar mass recovery}
We quantify stellar mass recovery using the difference between the SED-estimated and true stellar mass: $\Delta M_{\star} =\log_{10}(M_{\star,\rm{measured}}/{\rm{M_{\odot}}})-\log_{10}(M_{\star,\rm{true}}/\rm{M_{\odot}})$. We plot the distribution of $\Delta M_{\star}$ values for each SFH parametrisation and redshift in Figure \ref{fig:violin_plot}. To characterise the broad success of intrinsic stellar mass recovery, we then calculate the median and the standard deviation of $\Delta M_{\star}$. $\Delta M_{\star}=0$ would indicate no systematic offset in stellar mass estimates, $\Delta M_{\star}>0$ indicates an overestimated stellar mass, and  $\Delta M_{\star}<0$ indicates an underestimated stellar mass. We tabulate these quantities for each of the six SFH models, and for each of the six redshifts studied, in Table \ref{tab:mass_fits}. In general, stellar masses are recovered well when recovery is averaged over the whole population. There are some clear biases, though, with population-averaged stellar masses typically overestimated by SED fitting (median $\Delta M_{\star}>0$ for most redshifts and SFH parametrisations). Nevertheless, for all SFH parametrisations and redshifts, the median $\Delta M_{\star}$ is below $\sim0.4\,\rm{dex}$ (a factor of 2.5). The scatter in $\Delta M_{\star}$ is also fairly encouraging: this is below $0.3\,\rm{dex}$ (a factor of 2) for all SFH parametrisations and redshifts.\\
\indent Note that the summary values presented in Table \ref{tab:mass_fits} are averaged over the S{\sc phinx}$^{20}$ populations studied, which are themselves incomplete. The $(20\,\rm{cMpc})^{3}$ volume box was selected to probe a typical patch of the Universe in terms of halo mass distribution, and hence will not include the most massive galaxies at any epoch. In addition, the S{\sc phinx}$^{20}$ galaxies for which radiative transfer was run and post-processed spectra exist are limited to those with $\rm{SFR_{10}}>0.3\,\rm{M_{\odot}}yr^{-1}$. At the highest redshift studied, this results in samples of simulated galaxies that are preferentially low mass and high-sSFR. As we will show in Section \ref{sec:fitted_mass_trends}, stellar masses tend to be over-estimated for low mass, high-sSFR sources. This is reflected in the apparent trend of increasing over-estimation of stellar mass with increasing redshift seen in Table \ref{tab:mass_fits}. Due to the complex galaxy selection, these tabulated values should not be used as `corrections' to inferred SED-fitted values. In Sections \ref{sec:fitted_mass_trends}, we explore trends in stellar mass and SFR recovery as a function of several different intrinsic galaxy parameters.

\subsection{Trends in stellar mass recovery}\label{sec:fitted_mass_trends}
In Figure \ref{fig:mstar}, we show SED-fitted stellar mass, as well as the offset between fitted and true stellar mass, against true stellar mass, for the six different SFH parametrisations. There are clear stellar mass-dependent trends: at low stellar masses ($M_{\star}\lesssim10^{8}\,\rm{M_{\odot}}$), galaxy masses tend to be overestimated; at ($M_{\star}\lesssim10^{7}\,\rm{M_{\odot}}$), simulated galaxies can have their mass overestimated by almost an order of magnitude. This is broadly independent of redshift. In the range $M_{\star}\sim10^{8}-10^{9}\,\rm{M_{\odot}}$ (the mass range studied by \citealt{Narayanan2024}), masses are very well-recovered, typically to within a factor of two regardless of SFH parametrisation and redshift (see grey shaded regions). At the highest stellar masses ($M_{\star}\gtrsim10^{9}\,\rm{M_{\odot}}$), masses tend to be slightly underestimated, with slightly worse underestimations at $z=5$ compared to $z=6$. Importantly, these underestimations and overestimations are frequently larger than their associated uncertainties. The underestimation at high stellar masses is slightly less severe for the `Continuity' and `Bursty Continuity' models than for the other SFH parametrisations. The slope of $\Delta M_{\star}$ versus stellar mass varies only a little between SFH parametrisations and redshifts. It is, in general, very slightly flatter at higher redshifts (see orange and red lines marking the relations for $z=9$ and $z=10$, respectively). \\
\indent To investigate the cause of the trend in parameter recovery with stellar mass, we look for correlations between $\Delta M_{\star}$ and other galaxy properties. Since trends in $\Delta M_{\star}$ versus stellar mass are similar for the different SFH parametrisations, we adopt just a single parametrisation here (`Bursty Continuity'). We consider all redshifts studied (see labelled colored lines). As shown in Figure \ref{fig:mstar_vs_properties} (upper left-hand panel), there is a strong trend between $\Delta M_{\star}$ and specific star formation rate, calculated using the $10\,\rm{Myr}$-averaged SFR. A similar trend is seen between $\Delta M_{\star}$ and recent changes in star formation (parametrised by $\rm{SFR}_{10}/\rm{SFR}_{100}$; upper right-hand panel). Stellar masses are overestimated for galaxies with $\rm{SFR}_{10}/\rm{SFR}_{100}\gtrsim1$ (which corresponds to a `rising' SFH just prior to the epoch of observation; see also \citealt{Haskell2024}). We also calculate the offset between fitted and true stellar age, and plot this against $\Delta M_{\star}$ (lower right-hand panel). Where SED fits are overestimating the stellar mass, they also tend to be overestimating the stellar age, and thus the mass-to-light ratio. \\
\indent The strong correlation between recent SFR and overestimation of stellar mass hints at some underlying bias. One possibility is that this is caused by poorly fitted emission lines, as strong optical emission lines contaminate the redder JWST/PRIMER filters at these redshifts. At $z=5$, $\rm{H}\alpha$ is covered by the F356W, F410M, and F444W filters, and [O{\sc{iii}}]$\lambda\lambda4959,5007$ falls within the F277W filter. At $z=6$, [O{\sc{iii}}]$\lambda\lambda4959,5007$ falls within the F356W filter. At $z=7$, [O{\sc{iii}}]$\lambda\lambda4959,5007$ is covered by F356W, F410M and F444W. At $z=8$, [O{\sc{iii}}]$\lambda\lambda4959,5007$ is covered by F444W. At $z=9$, only the weaker [O{\sc{iii}}]$\lambda\lambda4363$ is covered by F444W. We find that $\Delta M_{\star}$ is indeed correlated with line equivalent width (EW; see lower left-hand panel of Figure \ref{fig:mstar_vs_properties}), except at $z=10$, where these lines are shifted out of the NIRCam filters. In Figure \ref{fig:EL_fits} (left-hand panels), we show correlations between offsets in SED-fitted and true line equivalent widths and offsets in stellar mass. Where line equivalent widths are underestimated, stellar masses are overestimated. We show two examples of SED fits that show this behaviour (right-hand panels). The upper right-hand panel shows the SED of a modelled $z=5$ galaxy, where strong $\rm{H}\alpha$ emission enters into the F356W, F410M, and F444W filters, and strong [O{\sc{iii}}]$\lambda\lambda4959,5007$ falls within the F277W filter. When a Double Power Law SFH is assumed (purple fit), {\small{BAGPIPES}} does not fit these strong emission lines, instead matching the photometry with an older stellar population and overestimating the stellar mass by $0.4\,\rm{dex}$. Adoption of the Bursty Continuity SFH (cyan fit) leads to a better match to the continuum and an accurate stellar mass estimate. The lower right-hand panel shows a modelled $z=8$ galaxy, where strong [O{\sc{iii}}]$\lambda\lambda4959,5007$ and $\rm{H}\beta$ lines fall within the F444W filter, and F356W, F410M and F444W are also boosted by the [O{\sc{iii}}]$\lambda4363$ line. In this case {\small{BAGPIPES}} fits a rising rest-optical SED, again underestimating line equivalent widths and overestimating the stellar mass regardless of SFH parametrisation. In summary, high line EWs drive red colors, and the SED fit favours fitting older stellar ages and higher mass-to-light ratios, driving overestimated stellar masses. \\
\indent Interestingly, we also see the opposite effect: where line equivalent widths are overestimated, stellar masses are slightly underestimated. In these cases, the modelled younger stellar populations drive stronger emission lines that approximately match the measured photometry (whereas in fact, the flux is driven by older stellar populations). The fitted continuum is less steeply rising in the rest-frame NIR than the true SED. Adding MIRI data should constrain this slope and increase the fitted stellar mass, minimising this bias.

\begin{figure*} 
\includegraphics[width=0.32\columnwidth]{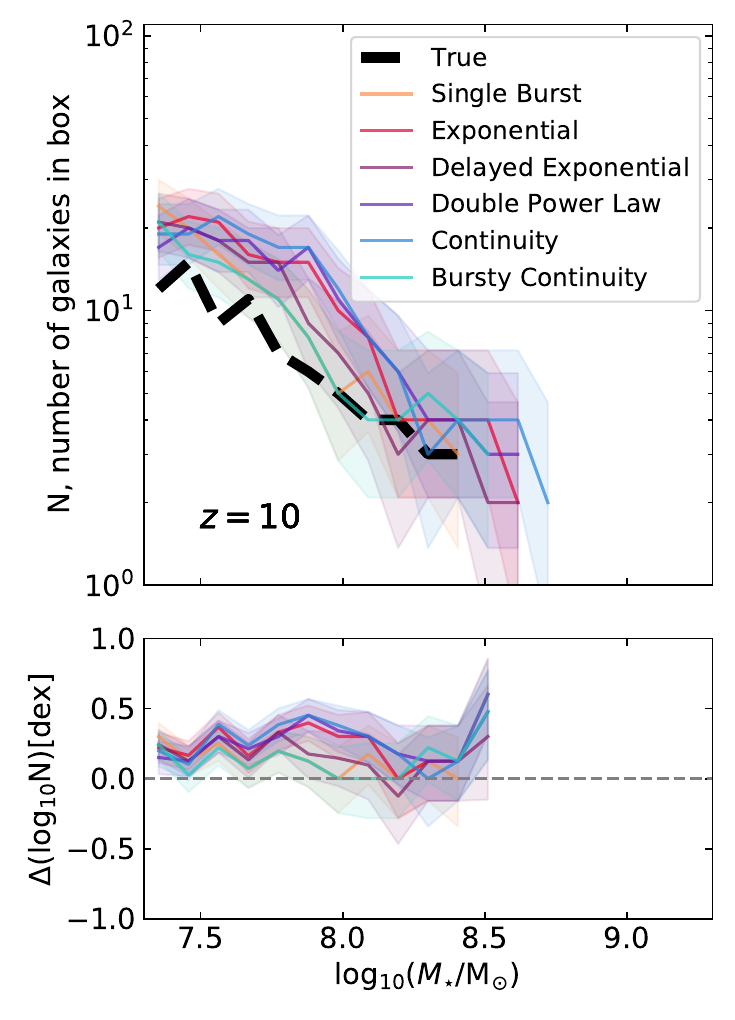}
\includegraphics[width=0.32\columnwidth]{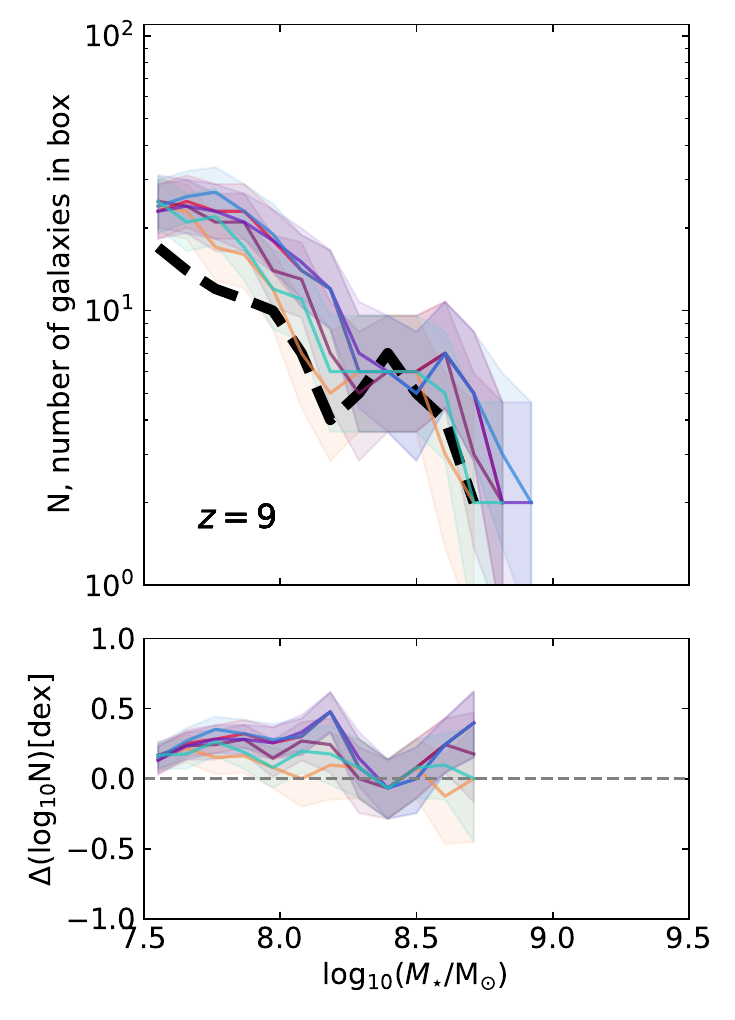}
\includegraphics[width=0.32\columnwidth]{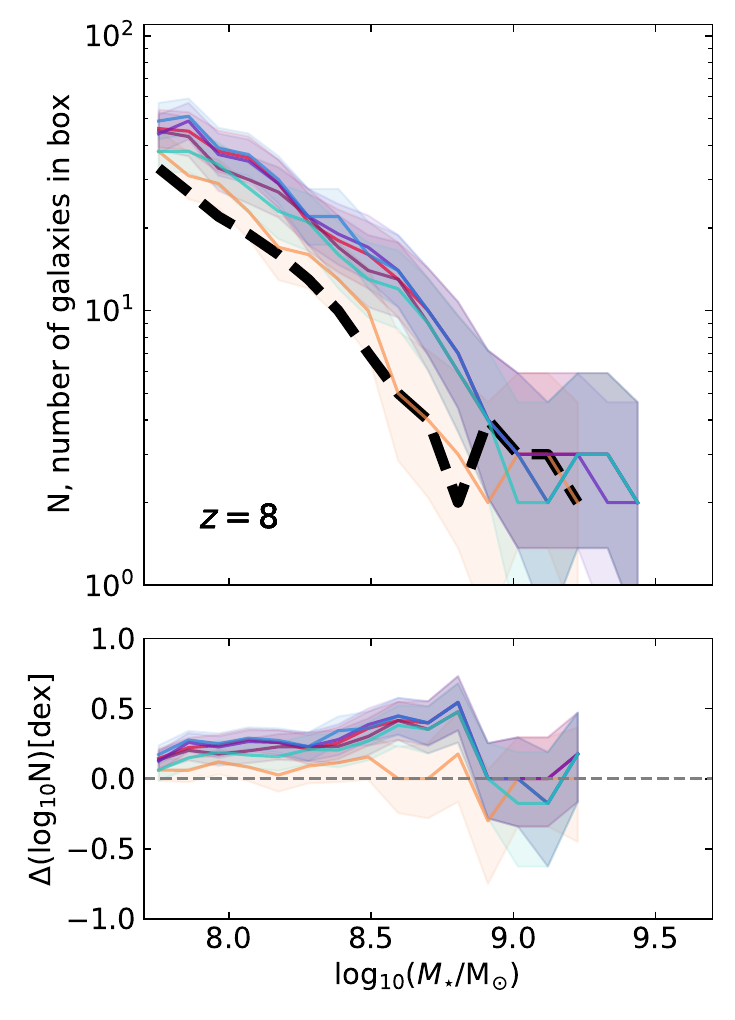}
\includegraphics[width=0.32\columnwidth]{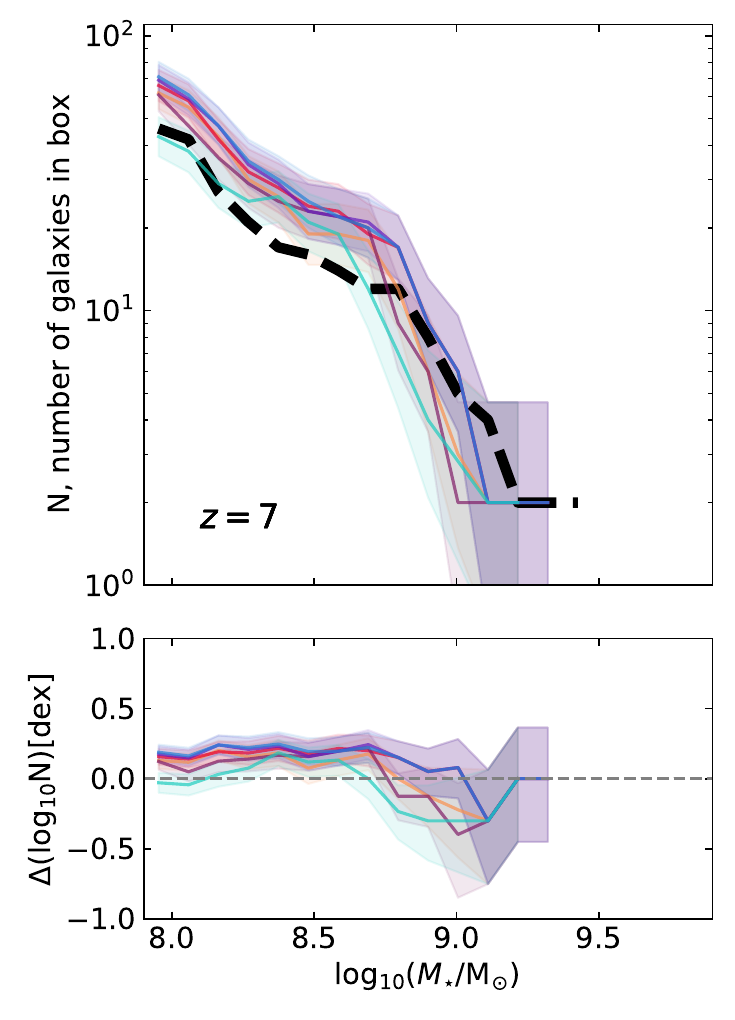}
\includegraphics[width=0.32\columnwidth]{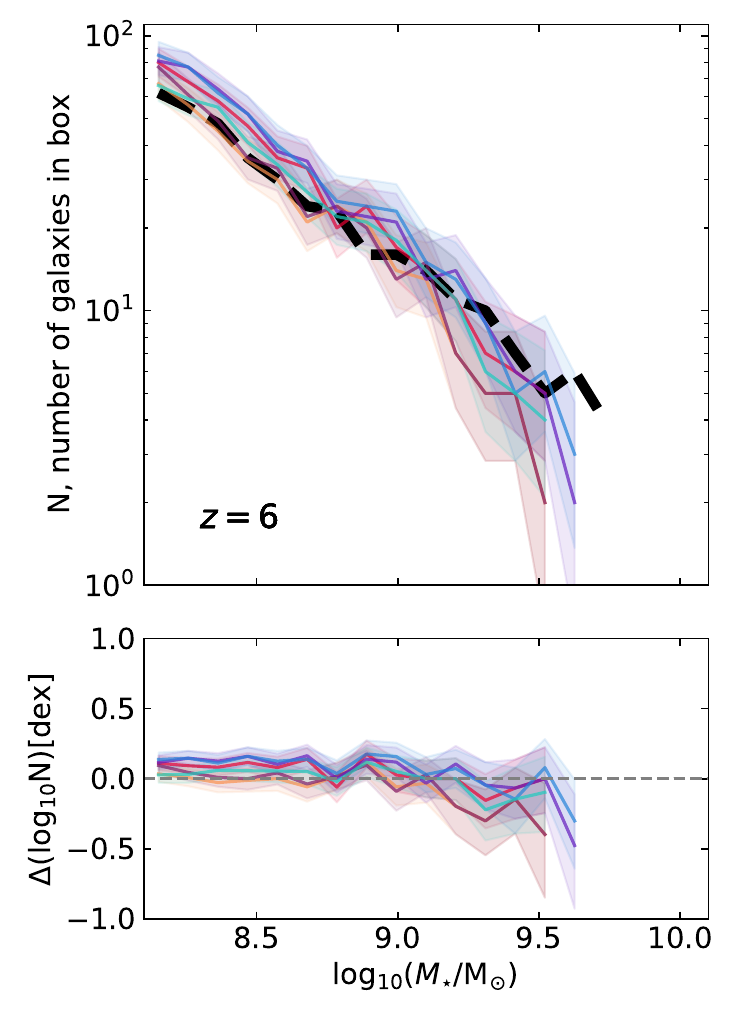}
\includegraphics[width=0.32\columnwidth]{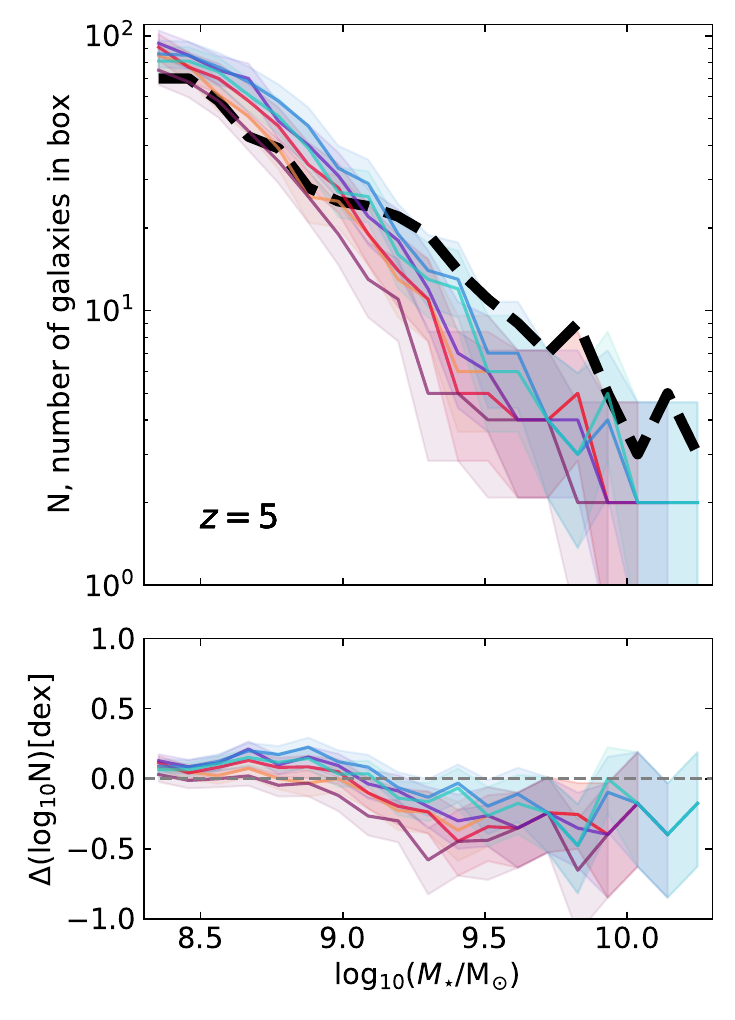}
\caption{Stellar mass functions at $z=10-5$, constructed using intrinsic simulated stellar masses (black dashed lines), as well as using masses inferred from SED fitting, with different assumed star formation histories (see colors shown in legend). The lower panels show $\log_{10}
\rm{N}_{\rm{inferred}}-\log_{10}\rm{N}_{\rm{true}}$. Uncertainties on the inferred functions are derived using Poisson errors. We do not include uncertainties on the simulated mass function as, although sometimes noisy, this is the `target' from the input data. The SED-inferred stellar mass functions are a different shape to the true functions, with more low-mass galaxies at high redshift ($z>7$) and fewer high-mass galaxies ($M_{\star}\gtrsim10^{9}\,\rm{M_{\odot}}$, seen here at $z\leq$7, where galaxies of these masses assemble). This `tilting' effect arises due to biases in stellar mass inference: masses are typically underestimated for high mass galaxies and overestimated for low-mass galaxies (which have high specific star formation rates to make it into the S{\sc phinx}$^{20}$ sample). Note that we do not display the lowest mass galaxies at each redshift, where number densities are heavily influenced by the SPHINX SFR selection. At these lowest stellar masses, the number of inferred galaxies is lower than the true number.\vspace{0.2cm}}
\label{fig:smfs}
\end{figure*}

\subsection{Impact on derived stellar mass functions}
Systematic biases in inferred galaxy parameters will have impacts on population summary functions constructed using SED fitting. Here, we explore the impact of biases in inferred stellar masses on stellar mass functions. First, we construct the intrinsic galaxy stellar mass function at each redshift (see black dashed lines in Figure \ref{fig:smfs}). Note that this is the stellar mass function of S{\sc phinx}$^{20}$ galaxies that are part of the data release, i.e. those for which $\rm{SFR}_{10}>0.3\,\rm{M_{\odot}}yr^{-1}$ and is hence incomplete. In some places, the stellar mass functions are also noisy, due to fairly small sample sizes. While not a prediction for the total stellar mass function (this would require folding in galaxies with $\rm{SFR}_{10}<0.3\,\rm{M_{\odot}}yr^{-1}$), it offers a benchmark against which we can compare the stellar mass function that would be constructed from the stellar masses inferred from SED fitting. In Figure \ref{fig:smfs}, we overplot the stellar mass function inferred using the different assumed star formation histories in different colors (see legend). Below each mass function, we plot the logged difference between inferred and true numbers of galaxies in each bin ($\log_{10}
\rm{N}_{\rm{inferred}}-\log_{10}\rm{N}_{\rm{true}}$). The SED-inferred stellar mass functions deviate quite substantially from the intrinsic functions. Regardless of SFH parametrisation, SED-inferred masses yield more low-mass galaxies (particularly at high redshift; $z>7$) and fewer high-mass galaxies ($M_{\star}\gtrsim10^{9}\,\rm{M_{\odot}}$, seen here at $z\leq$7, where galaxies of these masses begin to emerge). This can be attributed to the typical overestimation of stellar mass for low-mass, high-sSFR galaxies (which are those that make it into JWST-selected samples, given current survey depths) and underestimation for high-mass galaxies, as discussed in Section \ref{sec:fitted_mass_trends} and illustrated by Figure \ref{fig:mstar}. Overall, this effect leads to biases in inferred stellar mass functions. At $z=8-10$, the normalisation of the SMF is slightly overestimated (typically by $<0.5\,\rm{dex}$); this arises due to the shape of the SMF (more galaxies at low masses, scattering up to high masses). At lower redshifts, the inferred SMF is `tilted' steeper than the true SMF, as underestimation of the stellar masses of high-mass galaxies has a more prominent effect. The underestimation of the numbers of high mass galaxies is significant (up to $1\,\rm{dex}$).

\section{Discussion}\label{sec:discussion}
\subsection{Summary and comparison to previous work}
In this section, we summarise the key results of this Letter and compare to previous work. Using simulated galaxies where the `ground truth' stellar mass is known, with synthetic photometry generated using radiative transfer, we have tested a commonly-used SED fitting code with various assumptions regarding galaxy star formation history. Overall, stellar masses are recovered well for the simulated galaxies we study, which span $M_{\star}=10^{7}-10^{10}\,\rm{M_{\odot}}$ at $z=5-10$. $99,100,98,84,100\,\&\,100\%$ of stellar masses are recovered to within $0.5\,\rm{dex}$ at $z=5,6,7,8,9\,\&\,10$, respectively (for the `Bursty Continuity' model; see also Figure \ref{fig:violin_plot}); this is in stark contrast to the recent claim that `outshining' significantly hampers stellar mass inference at these redshifts \citep{Narayanan2024}. Importantly, we test six different models for the star formation history, four parametric and two non-parametric. There are only small differences in the fidelity of stellar mass inference between these models; this is possibly because only minimal cosmic time has passed by $z=5$, star formation histories are generally rising, and a very complex SFH is hence not required. Previous work has noted that exponential ($\tau$) and delayed-$\tau$ models typically do not perform well at low redshift \citep[e.g.][]{Lower2020}. This does not appear to be the case for our high-redshift galaxies, where both of these models are sufficiently flexible to fit a rising SFH toward the epoch of observation. \\
\indent We do observe significant trends in stellar mass recovery with galaxy stellar mass. The stellar masses of low stellar mass galaxies ($M_{\star}\lesssim10^{8}\,\rm{M_{\odot}}$; which in our sample are high-sSFR by selection) tend to be overestimated (by $\sim0.5\,\rm{dex}$ at $M_{\star}\sim10^{7}\,\rm{M_{\odot}}$), regardless of SFH parametrisation. Since our highest-redshift samples are comprised primarily of very low mass galaxies, the median stellar mass bias of the simulated sample appears more biased at high-$z$ as a result of this mass trend. Because of the S{\sc phinx}$^{20}$ sample selection ($\rm{SFR_{10}}>0.3\,\rm{M_{\odot}yr^{-1}}$), low mass galaxies in our sample are preferentially high-sSFR, and this may be driving the overestimation of stellar mass. Indeed, we showed that $\Delta M_{\star}$ correlates with $\rm{sSFR_{10}}$ and with the form of the recent SFH: galaxies with rising SFHs, with $\rm{SFR}(10\,\rm{Myr})>\rm{SFR}(100\,\rm{Myr})$, and with high line equivalent widths have more significantly overestimated stellar masses. Physically, this is driven by strong emission lines falling in the longer wavelength NIRCam bands driving redder colors. Regardless of SFH parametrisation, the SED fitting code (which uses CLOUDY modelling and our assumed BC03 stellar templates as input) is unable to fit the high EWs, and favours an older, more massive stellar population with higher mass-to-light ratio. Given that significant fractions of high-redshift galaxies appear to have extreme line emission (e.g. see \citealt{Boyett2023}), this is potentially a worry for high-redshift observational studies based only on broad-band JWST imaging. \\
\indent At higher stellar masses  ($M_{\star}\gtrsim10^{9}\,\rm{M_{\odot}}$, stellar masses tend to be underestimated (by up to $\sim0.5\,\rm{dex}$ at $M_{\star}\sim10^{10}\,\rm{M_{\odot}}$). Here, the SED fitting code fits the photometry with a younger, less massive stellar population, slightly overestimating emission line equivalent widths. We have demonstrated that this systematic bias, with stellar masses overestimated at low stellar masses and underestimated at high stellar masses, results in a bias in the stellar mass function constructed via SED fitted observations. At the highest redshifts studied ($z\geq8$), the normalisation of the mass function at $M_{\star}\lesssim10^{9}\,\rm{M_{\odot}}$ is overestimated by up to $0.5\,\rm{dex}$, while at higher masses (which are only probed by our simulations at $z\geq7$), numbers of galaxies are underestimated. Note that this is opposite to the impact of Eddington bias on the stellar mass function \cite[see also][]{Price2017}. If the underestimation of stellar mass at high stellar masses is due to poor emission line modelling or outshining of older stars by younger stellar populations, this may bias stellar mass estimates for star-forming rather than quiescent galaxies. If such an effect persists to lower redshifts ($z\lesssim4$), it could potentially result in over-estimation of fractions of quenched galaxies. 

\subsection{Future avenues}
\subsubsection{Broadening the photometric coverage}
In this study, we use the NIRCam bands that are part of the PRIMER survey, spanning $0.9-4.4\,\mu\rm{m}$. This results in different rest-frame wavelength coverage for the different redshifts studied: at $z=5$, $4.4\,\mu\rm{m}$ probes rest-frame $7300\,\angstrom$ emission, while at $z=10$, the same band probes $4000\,\angstrom$ emission. Several recent studies have shown that including MIRI photometry can improve stellar mass estimates for high-redshift galaxies. \cite{Song2023} found that stellar masses can be overestimated by up to $0.2\,\rm{dex}$ when rest-NIR data is not available, arguing that unbiased stellar mass estimates require data beyond rest-frame $1\,\mu\rm{m}$. \cite{Papovich2023} drew similar conclusions, noting that including MIRI $5.6\,\mu\rm{m}$ and $7.7\,\mu\rm{m}$ data reduced inferred stellar masses by $0.25\,\rm{dex}$ at $4<z<6$ and $0.37\,\rm{dex}$ at $6<z<9$, and that inferred SFRs were also lowered. This is because the MIRI data are able to identify strong nebular emission lines as the cause of rest-frame UV-to-optical SED reddening, ruling out alternate (incorrect) solutions of older stellar populations or dust attenuation. The improved constraints on the stellar mass provided by MIRI accordingly lower their estimated cosmic stellar mass density at $z=9$ by an order of magnitude compared to pre-JWST studies. The addition of NIRCam medium bands would also enable better identification of emission line contamination and constraints on line-free regions on the SED (see Appendix \ref{sec:medium_bands}). Future work should extend wavelength coverage and also test different filter combinations.

\subsubsection{Other systematics and assumptions}
Photometric redshift estimation is complex \cite[see e.g.][]{Salvato2019,Newman2022} and beyond the scope of this study. Following \cite{Narayanan2024}, we have considered the `best-case' scenario in which the correct redshift is known. Future work will include studying the impact of uncertain redshifts on inferred galaxy physical properties.\\
\indent More broadly, as discussed in the Introduction, many choices besides SFH parametrisation are made in the process of SED fitting. These include the selection of the stellar IMF, the treatment of dust, as well as inclusion or not of prescriptions for AGN emission. In this work, we adopt standard choices used in observational work (a \citealt{Kroupa2002} IMF; a flexible dust model that enables a variety of attenuation curves to be fitted; no AGN emission). Substantial discrepancies between fitting assumptions and truth (for example very different IMFs) will lead to further systematic errors in parameter inference, and future work should explore these. Given the significant role of emission lines in driving biased stellar mass measurements, it will be important to extend this work to test different stellar population synthesis templates and methods of emission line modelling.

\section{Conclusions}\label{sec:conclusions}
In this Letter, we have made use of the S{\sc phinx}$^{20}$ cosmological radiation hydrodynamics simulation to study the inference of stellar mass at high-redshift with JWST photometry. Our work expands upon previous efforts that employ forward-modelled simulation data, where the `ground truth' galaxy physical properties are known, to test the SED codes commonly used to fit observational data \citep[e.g.][]{Hayward2014,Haskell2023,Haskell2024}. We use a standard implementation of the {\sc bagpipes} code to fit the modelled NIRCam photometry of simulated galaxies at $z=5-10$. We use the following NIRCam bands: F090W, F115W, F150W, F200W, F277W, F356W, F410M and F444W, and assume that a signal-to-noise ratio of $10$ is achieved in each band. Regardless of our choice of star formation history parametrisation (single burst, exponential, delayed exponential, double power law, continuity, bursty continuity), galaxy stellar masses are fairly well-recovered (generally to within a factor of $2$ in the range $M_{\star}\sim10^{7.5}-10^{9.5}\,\rm{M_{\odot}}$). This success is in contrast to the recent work of \cite{Narayanan2024}, who performed similar tests for modelled galaxies at $z=7$ and argued that stellar masses were very poorly recovered, even with the extended photometric coverage they assumed. Our results should be generally encouraging for prospects of deriving the stellar masses of high-redshift galaxies from imaging surveys. \\
\indent Nevertheless, we observe some significant biases in stellar mass recovery. The stellar masses of low-mass galaxies, which to make it into our selection have high specific star formation rates, tend to be overestimated (by $\sim0.5\,\rm{dex}$ at $M_{\star}\sim10^{7}\,\rm{M_{\odot}}$). We demonstrate that this is driven by strong emission lines driving redder colors. Our SED modelling setup is unable to fit the high line equivalent widths, and instead favours older, more massive stellar populations with higher mass-to-light ratios. The stellar masses of more massive galaxies tend to be slightly underestimated (by up to $\sim0.5\,\rm{dex}$ at $M_{\star}\sim10^{10}\,\rm{M_{\odot}}$). This systematic bias is potentially worrying for high-redshift observational studies based only on broad-band JWST imaging, \citep[see also][]{Boyett2023,Song2023,Papovich2023}, and could potentially impact the shape and normalisation of the inferred stellar mass function. Planned future work will explore additional uncertainties involved in fitting galaxy redshifts, as well as the impact of redshift failures on stellar masses (here, we fixed redshifts at the true values, following \citealt{Narayanan2024}). We will also investigate the impact of other fitting choices on inferred stellar masses, focusing in particular on the choice of stellar population synthesis models. \\ 

\section*{Acknowledgements}
\noindent We thank the anonymous referee for helpful suggestions. RKC was funded by support for program \#02321, provided by NASA through a grant from the Space Telescope Science Institute, which is operated by the Association of Universities for Research in Astronomy, Inc., under NASA contract NAS 5-03127. RKC is grateful for support from the Leverhulme Trust via the Leverhulme Early Career Fellowship. The Flatiron Institute is supported by the Simons Foundation. PNB is grateful for support from the UK STFC via grant ST/V000594/1. RKC acknowledges Ross McLure and Desika Narayanan for helpful conversations. We are grateful to the S{\sc phinx} team for making their simulation data publicly available.\\
\section*{Data availability}
\noindent Version 1 of the S{\sc phinx}$^{20}$ data release is publicly available to download from https://github.com/HarleyKatz/SPHINX-20-data.

\bibliographystyle{aasjournal}
\bibliography{main}

\appendix
\section{The impact of adding full JWST medium band coverage}\label{sec:medium_bands}
\begin{figure*}
\includegraphics[width=0.33\columnwidth]{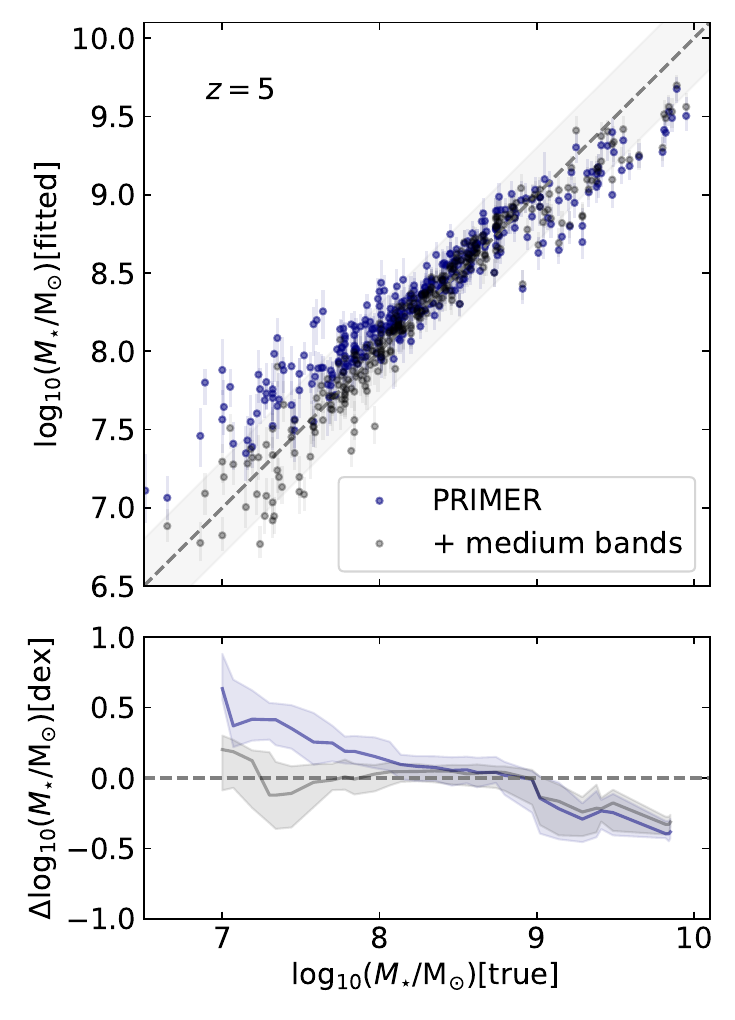}
\includegraphics[width=0.33\columnwidth]{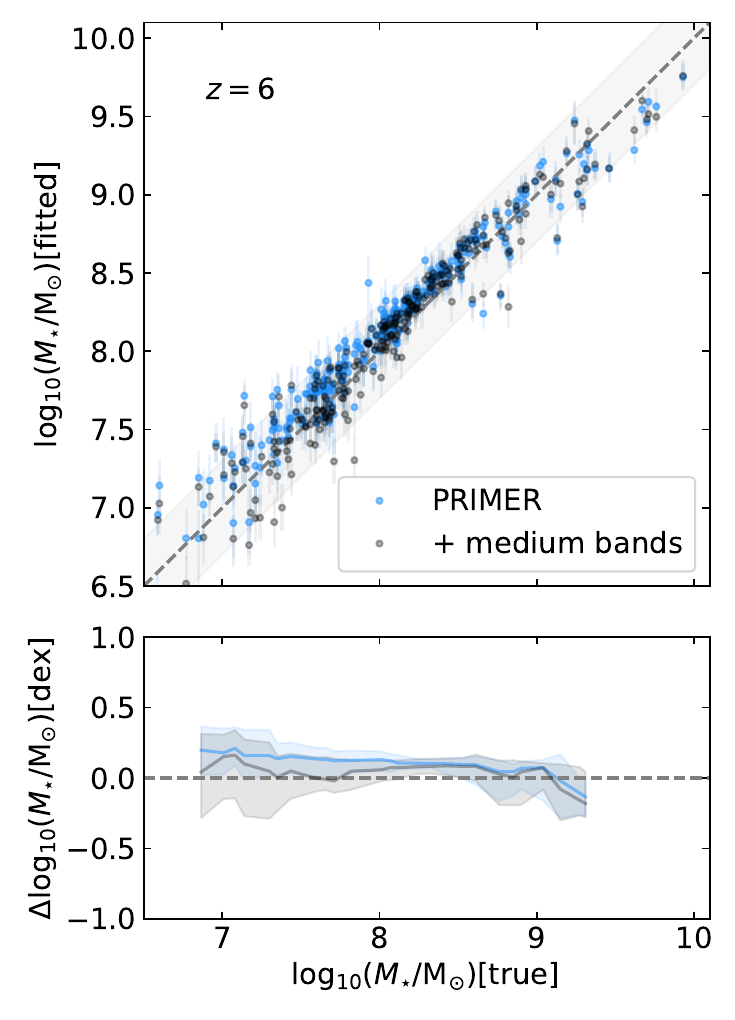}
\includegraphics[width=0.33\columnwidth]{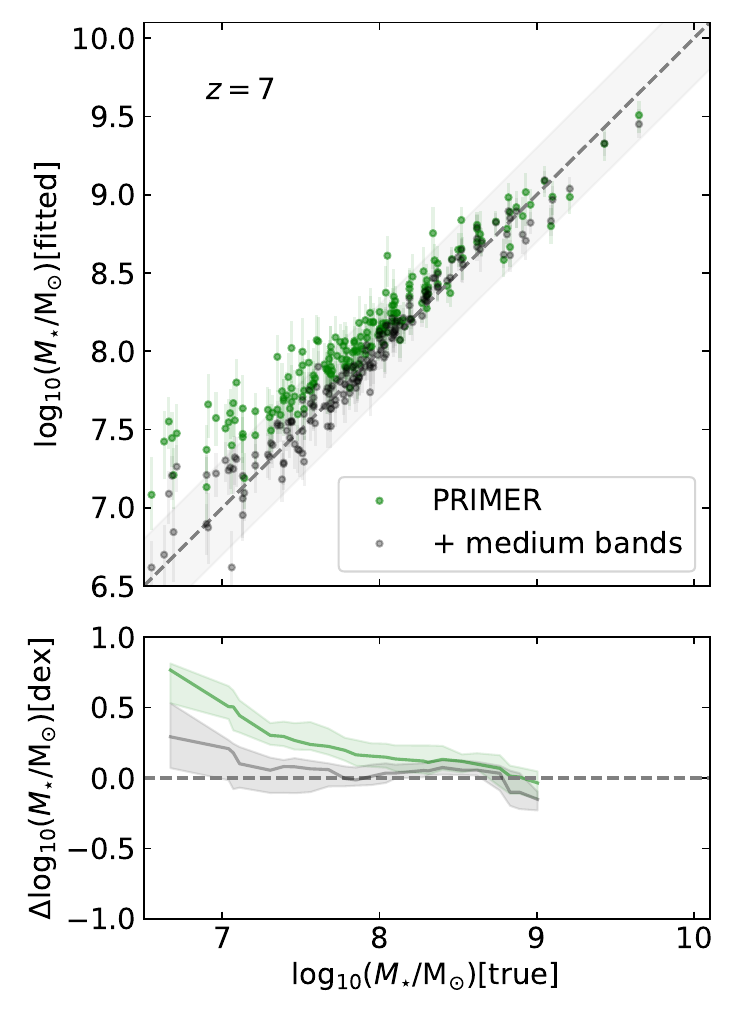}
\includegraphics[width=0.33\columnwidth]{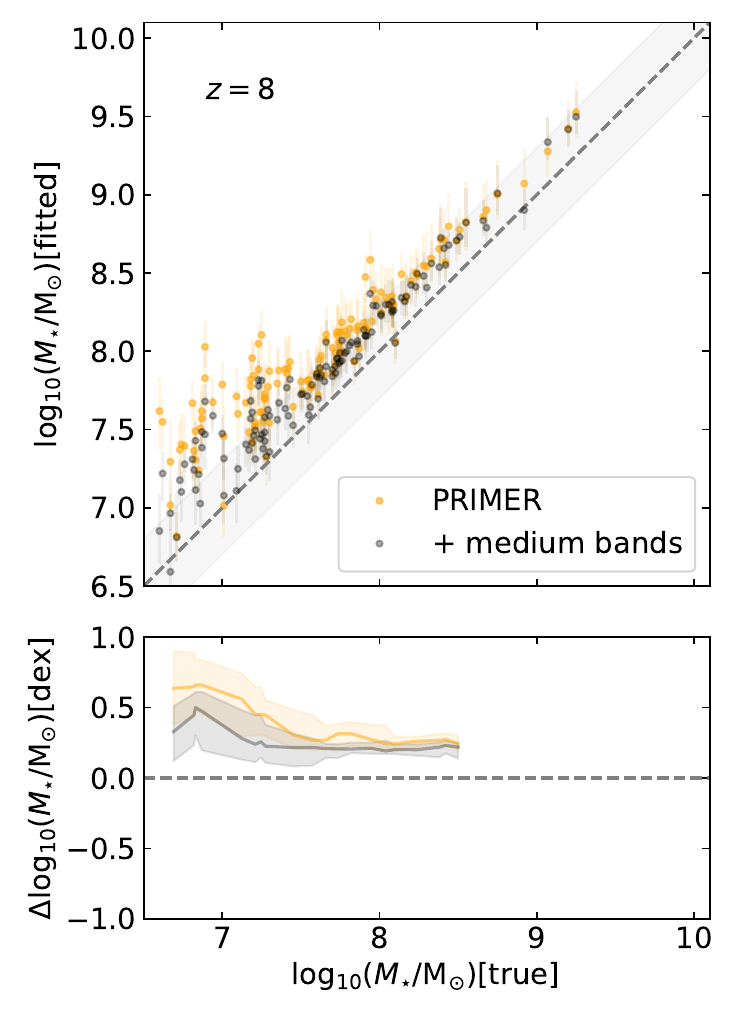}
\includegraphics[width=0.33\columnwidth]{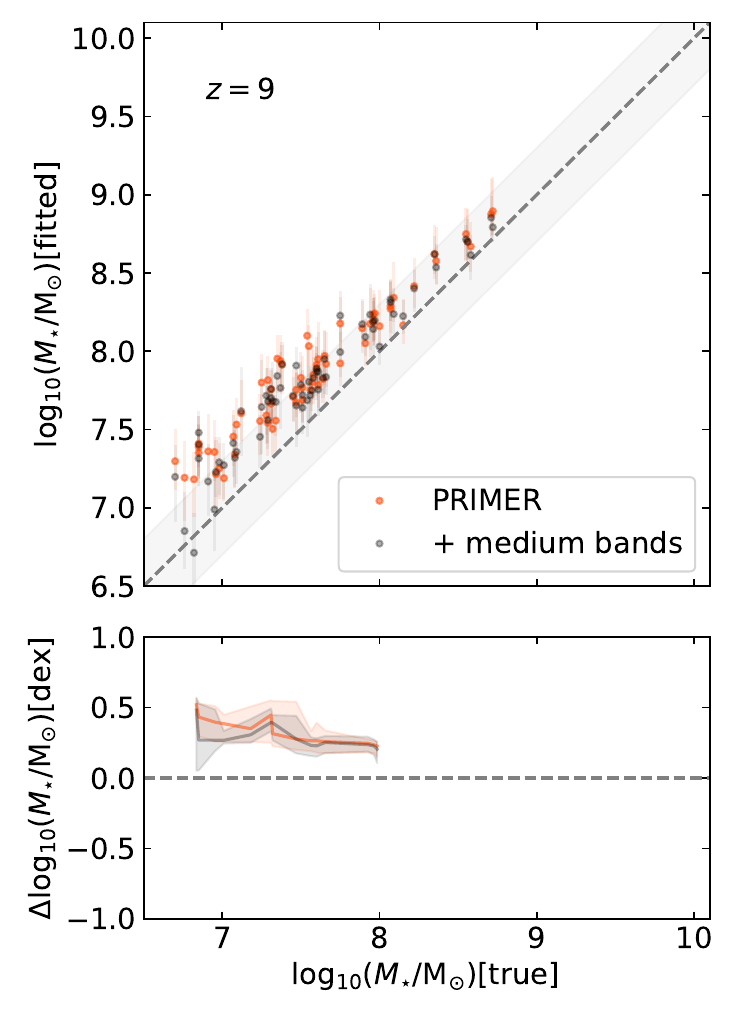}
\includegraphics[width=0.33\columnwidth]{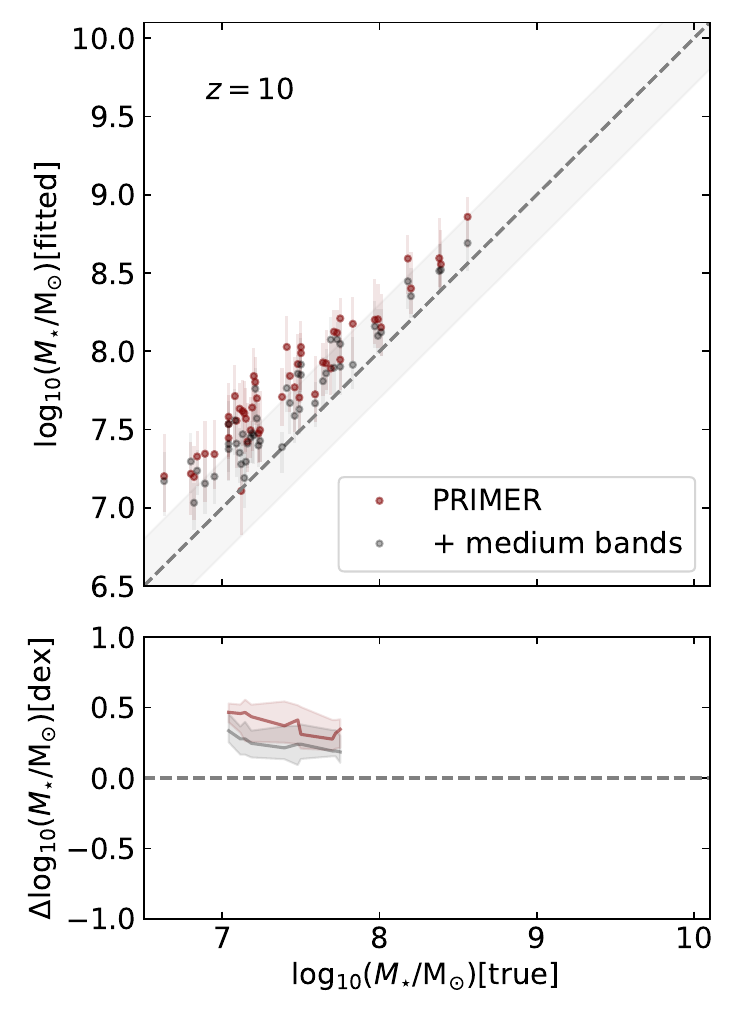}
\caption{The impact of adding the remaining JWST medium bands (F140M, F210M, F250M, F300M, F335M, and F480M) to the existing set of PRIMER filters on the inferred stellar masses, at each redshift (different panels). Stellar masses derived from SED fits including all broad and medium bands are shown in black on each panel, and the original fits (using just the PRIMER filter coverage) are shown in color. All fits use the Double Power Law SFH parametrisation. Particular improvement is seen for low-mass galaxies at $z=5$ (where the addition of the F250M, F335M and F480M data enable continuum constraints), and at $z=7$ and $z=8$ (where the F480M measurement provides an important constraint against a rising continuum).\vspace{0.2cm}}
\label{fig:medium_bands}
\end{figure*}

\noindent In this study, we restricted photometric coverage to the eight NIRCam filters used in existing PRIMER coverage of COSMOS and UDS. Here, we refit SEDs for all the simulated galaxies with the addition of the remaining NIRCam medium bands (F140M, F210M, F250M, F300M, F335M, and F480M), and quantify the impact on stellar mass recovery. For simplicity, we use only the Double Power Law SFH parametrisation. We present the refitted stellar masses (black points), alongside original fitted masses (colored points), in Figure \ref{fig:medium_bands}. There is a significant improvement in stellar mass recovery. The percentage of masses recovered to within $0.5\,\rm{dex}$ increases from $94\%$ to $99\%$ at $z=5$, from $91\%$ to $99\%$ at $z=7$, from $76\%$ to $91\%$ at $z=8$, from $85\%$ to $95\%$ at $z=9$ and from $84\%$ to $96\%$ at $z=10$. At $z=6$, there is no change on the excellent recovery ($99\%$). At $z=5$, the addition of the F250M, F335M and F480M filters, which are not contaminated by emission lines, enables stronger constraints on the continuum. At higher redshifts, the F480M filter provides a very important anchor against a rising continuum.

\end{document}